\documentclass[a4paper,fleqn]{cas-dc}

\def\tsc#1{\csdef{#1}{\textsc{\lowercase{#1}}\xspace}}
\tsc{WGM}
\tsc{QE}
\tsc{EP}
\tsc{PMS}
\tsc{BEC}
\tsc{DE}
\usepackage{amsmath,amssymb,amsfonts, amsthm}
\usepackage{algorithmic}
\usepackage{graphicx}
\usepackage{textcomp}
\usepackage{acronym}
\usepackage{booktabs}
\usepackage[numbers,sort&compress]{natbib}
\usepackage{array}
\usepackage{tabularx}
\usepackage{graphicx}
\usepackage{makecell}
\usepackage{ragged2e}

\usepackage{xcolor}
\usepackage{comment}
\usepackage{pifont}
\usepackage{subcaption} 
\usepackage[font=small]{caption}

\acrodef{aiits}[AIITS]{All-Island Irish Transmission System}
\acrodef{avr}[AVR]{automatic voltage regulator}
\acrodef{coi}[COI]{center-of-inertia}
\acrodef{dae}[DAE]{differential-algebraic equation}
\acrodef{ddae}[DDAE]{delay differential-algebraic equation}
\acrodef{der}[DER]{distributed energy resource}
\acrodef{fr}[FR]{frequency regulation}
\acrodef{pss}[PSS]{power system stabilizer}
\acrodef{sg}[SG]{synchronous generator}
\acrodef{sssa}[SSSA]{small-signal stability analysis}
\acrodef{wams}[WAMS]{wide area measurement system}
\acrodef{fem}[FEM]{forward euler method}
\acrodef{tm}[TM]{trapezoidal method}
\acrodef{ks}[KS]{Krylov-Schur}

\newcommand{\T}{^{\intercal}}

\newcommand{\bfg}[1]{\boldsymbol{#1}}
\newcommand{\bfb}[1]{\boldsymbol{\rm #1}}
\newcommand{\jj}{\jmath}
\newcommand{\nx}{n}
\newcommand{\ny}{m}
\newcommand{\nxy}{r}
\newcommand{\xys}{\ensuremath{\mbox{$\bfb {x}$}}}

\newcommand{\jac}[2]{\bfg{{#1}}_{\hspace{-0.2mm}#2}}
\newcommand{\del}{j}
\newcommand{\Cs}{\bfb C}
\newcommand{\Sn}{\bfb S}
\newcommand{\Csd}{\bfb C_{\scriptscriptstyle D}}
\newcommand{\Snd}{\bfb S_{\scriptscriptstyle D}}

\definecolor{blue1}{RGB}{33,92,175}

\def\BibTeX{{\rm B\kern-.05em{\sc i\kern-.025em b}\kern-.08em
    T\kern-.1667em\lower.7ex\hbox{E}\kern-.125emX}}

\begin{document}

\let\WriteBookmarks\relax
\def\floatpagepagefraction{1}
\def\textpagefraction{.001}
\shorttitle{}
\shortauthors{A. Bouterakos et~al.}

\title [mode = title]{Sparse Continuation-Based Eigenvalue Tracking for Power System DDAEs}
\tnotemark[1]

\tnotetext[1]{This work is supported by the CETPartnership Joint Call 2024 under project NU-ACTIS, co-funded by the European Commission (grant no. 101069750).  The participation of University College Dublin is funded by the Sustainable Energy Authority of Ireland (ID 24/RDD/1390).}

\author[1]{A. Bouterakos}[type=editor,
                        auid=000,bioid=1,
                        orcid=0009-0008-9900-4648]
\ead{andreas.bouterakos@ucdconnect.ie}

\credit{Writing – review \& editing, Writing – original draft, Visualization, Validation, Software, Methodology, Investigation, Formal analysis, Data curation, Conceptualization}

\affiliation[1]{organization={School of Electrical and Electronic Engineering, University College Dublin},
                addressline={Belfield}, 
                city={Dublin},
                postcode={D04 V1W8},
                country={Ireland}}

\author[1]{G. Tzounas}[type=editor,
                        auid=000,bioid=1,
                        orcid=0000-0002-1464-3600]
\cormark[1]
\ead{georgios.tzounas@ucd.ie}

\credit{Writing – review \& editing, Writing – original draft, Visualization,Validation, Supervision, Software, Resources, Project administration, Methodology, Investigation, Funding acquisition, Formal analysis, Data curation, Conceptualization}
\cortext[cor1]{Corresponding author}
\begin{abstract}
In this paper, we formulate a continuation method for tracking eigenvalue trajectories in power system models with time-delayed measurement and control signals. Such delays are known to weaken damping and reduce stability margins if not properly accounted for in stability analysis and control design. The proposed formulation follows selected eigenpairs directly from sparse \ac{ddae} models with one or multiple delayed variables,
avoiding repeated eigensolutions as system conditions or parameters vary.
The continuation parameter can represent system and control parameters, constant delay magnitudes, or parameters governing non-constant wide-area measurement system (WAMS) delays.  The proposed method is validated on a modified IEEE 39-bus system and on a real-world-scale dynamic model of the Irish transmission network, demonstrating accurate tracking of critical eigenvalue trajectories and clear computational advantages over repeated eigensolution-based analysis.
\end{abstract}


\maketitle

\section{Introduction}

\subsection{Motivation}\label{motivation}

Time delays in power system control loops arise from measurement, data processing, and communication. These delays, which typically range from a few milliseconds to hundreds of milliseconds, can lead to sustained oscillations and compromise stability
\cite{liu2018stability,zografopoulos2023distributed, concepcion2017effects, stahlhut2008latency,tzounas2021damping}.
These effects can be assessed through \ac{sssa}, where eigenvalue analysis plays a central role \cite{book:eigenvalue}.
However, in the presence of delays, eigenvalue analysis typically relies on spectral approximations, making repeated eigenvalue computations expensive and tracking eigenvalue trajectories across parameter variations nontrivial, particularly in large-scale settings.  Addressing the problem of tracking eigenvalue trajectories in power system models with time delays is the main goal of this paper.

\subsection{Literature Review}

Power system dynamics in short-term stability analysis are commonly modeled through a set of \acp{dae} \cite{kundur:94}.  Introducing time delays transforms the model into a set of \acfp{ddae}.  In \ac{sssa}, the stability of linearized \acp{dae} and \acp{ddae} is determined by their eigenvalues, defined as the roots of the associated characteristic equation.  For \acp{ddae}, this equation is transcendental and gives rise to an infinite spectrum \cite{book:eigenvalue}. 
Consequently, a finite-dimensional approximation is required and its accuracy depends on the specific approximation employed.
Rational techniques such as Pad{\'e} polynomials, which are commonly used in commercial software implementations, approximate the 
transcendental terms themselves, with the accuracy of the resulting eigenvalues depending on factors such as the approximation order and delay magnitude.  
Spectral discretization provides a more accurate alternative by approximating the underlying infinite-dimensional eigenvalue problem directly \cite{milano2015small}.
\color{black}
The finite-dimensional linear eigenvalue problem that arises from the above techniques can then be solved using standard numerical algorithms \cite{tzounas2020comparison}.
These include dense-matrix methods, such as QR or QZ, which are suitable for small to medium-sized systems, 
but become computationally prohibitive when repeated eigendecompositions are required as the problem scales;
and sparse techniques, such as Krylov methods, which can obtain selected parts of the spectrum in large systems, but rely on user-defined spectral transformations and target regions. This makes their repeated use nontrivial when system parameters vary and eigenvalue trajectories must be followed.
\color{black}

This difficulty motivates methods that follow eigenvalues as continuous functions of the varying parameters, rather than recomputing and re-identifying them independently at each parameter value. Such a framework 
is provided by continuation methods. The core idea is to incrementally update the solution of a parameter-dependent problem using previously computed values.  In the context of eigenvalue tracking, this solution consists of a subset of the system's eigenpairs, i.e., eigenvalues with their associated eigenvectors. 
For systems without delays,
\cite{dieci2001continuation} proposes a continuation-based algorithm that employs Newton's method to follow eigenvalue trajectories, highlighting the benefit of including a predictor step before each iteration.
The authors in \cite{LIBERDA2003517} combine this approach with a Cayley transform that updates the dimension of the considered subspace at every iteration, focusing the computation on the modes most critical for stability.  
To evaluate power system stability margins, \cite{Wen_2006} derives eigenvalue and eigenvector sensitivities with respect to varying parameters and traces their trajectories through numerical integration.  In all these works, time delays are neglected.

Among the few works that do consider delays, existing formulations are limited to constant delay models \cite{mou2021delayed,li2017eigenvalue}. Both formulations assume a fixed steady-state solution and and therefore cannot capture parameter variations that alter the system power flow, such as changes in active and reactive load powers.  Moreover, 
the formulation in \cite{li2017eigenvalue} uses the delay magnitude as the continuation parameter, restricting the range of parameter variations that can be studied.
Finally, formulations based on dense Sylvester-type equations become computationally prohibitive as the problem scales, are therefore unsuitable for the study of large systems.
Addressing these limitations is the focus of this paper.
\color{black}

\subsection{Contribution}

Motivated by the limitations identified above, this paper develops a continuation-based eigenvalue tracking formulation for large-scale power system \acp{ddae}. The main advances compared with existing techniques are as follows:

\begin{itemize}
\item The formulation works directly with the sparse \ac{ddae} eigenvalue problem. This avoids dense matrix equations, whose computational cost becomes prohibitive as the problem scales.

\item 
The continuation parameter is not limited to the delay magnitude. It can represent system parameters, control parameters, delay magnitudes, or combinations of them, 
including parameters that alter the steady-state operating point, such as active and reactive load powers.
\color{black}

\item The formulation can handle multiple delayed variables and can account for non-constant \ac{wams} delays, including stochastic latency effects and data packet dropouts. This supports a more complete assessment of wide-area controllers.

\end{itemize}

\subsection{Paper Organization}

The remainder of the paper is organized as follows.  Section~\ref{sec:sssa} introduces preliminaries on \ac{sssa}.  Section~\ref{sec:eigtrack} discusses the formulation and implementation aspects of the proposed eigenvalue tracking approach. Section~\ref{sec:case_study} presents simulation results, based on the modified 39-bus system and a real-world-scale model of the Irish transmission system, demonstrating the accuracy and computational performance of 
the method. Finally,  Section~\ref{sec:conclusions} draws conclusions and outlines directions for future work.

\section{Small-Signal Stability Analysis}\label{sec:sssa}

\subsection{DDAE Power System Model}

Short-term power system stability in the presence of delays can be studied through a set of nonlinear
\acp{ddae} in 
semi-implicit\footnote{
The semi-implicit form is often adopted because it can offer certain practical advantages
such as increased sparsity as well as flexibility in handling state and algebraic variables, see \cite{book:eigenvalue}.} 
form
\cite{book:eigenvalue}
\begin{align}
\label{eq:ret_sidae}
  \begin{bmatrix}
    \bfg T & \bfg 0_{\nx,\ny} \\
    \bfg R & \bfg 0_{\ny,\ny}
  \end{bmatrix}
  \begin{bmatrix}
    \dot{\bfg x}(t) \\
    \dot{\bfg y}(t)
  \end{bmatrix}  &=
  \begin{bmatrix}
        \bfg f( \bfg x, \bfg y, \bfg x_{d}, \bfg y_{d} ) \\
        \bfg g( \bfg x, \bfg y, \bfg x_{d}, \bfg y_{d} )
  \end{bmatrix} 
\end{align}
where $\bfg x = \bfg x(t): [0,\infty) \rightarrow \mathbb{R}^{\nx}$ are the states of dynamic devices such as generators, dynamic loads, and controllers; $\bfg y=\bfg y(t): [0,\infty) \rightarrow \mathbb{R}^{\ny}$ are algebraic variables associated with network equations and auxiliary control setpoints; $\bfg f$ and $\bfg g$ are nonlinear functions; $\bfg T \in \mathbb{R}^{\nx \times \nx}$, $\bfg R\in \mathbb{R}^{\ny \times \nx}$;
and $\bfg 0_{\nx,\ny}$ denotes the $\nx \times \ny$ zero matrix.
The delayed state and algebraic variables are denoted as $\bfg x_d$ and $\bfg y_d$, with:
\begin{equation}
\begin{aligned}
\label{eq:xy_delayed}
    \bfg x_d(t) &= 
    \{ \bfg x(t-\tau_1), \bfg x(t-\tau_2), \ldots, \bfg x(t-\tau_\mu) \} \\
     \bfg y_d(t) &= 
    \{ \bfg y(t-\tau_1), \bfg y(t-\tau_2), \ldots, \bfg y(t-\tau_\mu) \}
\end{aligned}
\end{equation}
where $\tau_\del > 0$, $\del=1,2,\ldots,\mu$, denotes the $\del$-th delay and $\mu$ is the total number of delays.  For brevity, we denote the $\del$-th delayed variable as:
\begin{equation}
\begin{aligned}
\label{eq:tau_notation}
\bfg x^\del_d &= \bfg x(t-\tau_\del) \ , 
\\
\bfg y^\del_d &= \bfg y(t-\tau_\del) \ ,
\end{aligned}
\end{equation}
where $\del=1,2,\ldots,\mu$.

\subsection{Eigenvalue Analysis}

To study how the system responds to parameter variations, we introduce a scalar continuation parameter $p \in \mathbb{R}$,
used to parametrize selected system quantities of interest.
System parameters in our formulation are treated as functions of $p$, which allows tracking along any prescribed path and thus accommodates simultaneous variations of several parameters.
\color{black}
Then, \eqref{eq:ret_sidae} is rewritten as follows:
\begin{align}
\label{eq:ret_sidae:p}
  \begin{bmatrix}
    \bfg T(p) & \bfg 0_{\nx,\ny} \\
    \bfg R(p) & \bfg 0_{\ny,\ny}
    \end{bmatrix}
      \begin{bmatrix}
       \dot{\bfg x} \\ \dot{\bfg y}
    \end{bmatrix}  &=
    \begin{bmatrix}
      \bfg f( \bfg x, \bfg y, \bfg x_d, \bfg y_d, p ) \\
      \bfg g( \bfg x, \bfg y, \bfg x_d, \bfg y_d, p )
    \end{bmatrix} 
\end{align}
 
An equilibrium $(\bfg x_o, \bfg y_o):= [\bfg x_o\T, \bfg y_o\T]$
($\T$ indicating the transpose) of \eqref{eq:ret_sidae:p} can be defined assuming the system has remained at rest for a time at least equal to the maximum delay.  Considering small disturbances, 
\eqref{eq:ret_sidae:p} can be linearized around the equilibrium, as follows:
  \begin{align}
  \nonumber
  \begin{bmatrix}
    \bfg T(p) & \bfg 0_{\nx,\ny} \\
    \bfg R(p) & \bfg 0_{\ny,\ny}
    \end{bmatrix}
      \begin{bmatrix}
      \Delta \dot{\bfg x} \\
      \Delta \dot{\bfg y}
    \end{bmatrix}
    =
    \begin{bmatrix}
     \jac{f}{x}(p) & \jac{f}{y}(p)  \\
     \jac{g}{x}(p) & \jac{g}{y}(p) 
    \end{bmatrix} 
    \begin{bmatrix}
      \Delta {\bfg x} \\
      \Delta {\bfg y}
    \end{bmatrix}\\
    + \sum_{\del=1}^{\mu}(
    \begin{bmatrix}
     \jac{f}{x,\del}(p) & \jac{f}{y,\del}(p)  \\
     \jac{g}{x,\del}(p) & \jac{g}{y,\del}(p) 
    \end{bmatrix} 
    \begin{bmatrix}
      \Delta {\bfg x^\del_d} \\
      \Delta {\bfg y^\del_d}
    \end{bmatrix})
    \label{eq:ret_sidae:lin}
\end{align}
where $\Delta \bfg x = \bfg x - \bfg x_o$, $\Delta \bfg y = \bfg y - \bfg y_o$; $\jac{f}{x}$, $\jac{f}{y}$, $\jac{g}{x}$, $\jac{g}{y}$ and $\jac{f}{x,\del}$, $\jac{f}{y,\del}$, $\jac{g}{x,\del}$, $\jac{g}{y,\del}$ are the delay-free and delayed Jacobians, respectively, evaluated at $(\bfg x_o, \bfg y_o)$.  Equation \eqref{eq:ret_sidae:lin} is of the form: 
\begin{align}
\label{eq:ret_singps}
  \bfb{E} (p) 
  \, \dot{\xys} = \bfb A_0(p) \, \xys + \sum_{\del=1}^{\mu}\bfb A_\del(p) \, \xys^\del_d 
\end{align}
where 
\begin{align*}
\xys &=(\Delta \bfg x, \Delta \bfg y) \ , \\ 
\xys^\del_d &=(\Delta \bfg x^\del_d, \Delta \bfg y^\del_d)  \ ,
\end{align*}
%
and
\begin{equation}
\label{eq:ret_matrices:sidae}
 \begin{aligned}
  \bfb{E}(p) &\equiv 
  \begin{bmatrix}
    \bfg T(p) & \bfg 0_{\nx,\ny} \\
    \bfg R(p) & \bfg 0_{\ny,\ny} \\
  \end{bmatrix}
   \, , \
  \bfb{A}_0(p) \equiv 
  \begin{bmatrix}
   \jac{f}{x}(p) & \jac{f}{y}(p) \\
   \jac{g}{x}(p) & \jac{g}{y}(p) \\
  \end{bmatrix} 
   \\ 
  \bfb{A}_\del(p) &\equiv 
  \begin{bmatrix}
   \jac{f}{x,\del}(p) & \jac{f}{y,\del}(p) \\
   \jac{g}{x,\del}(p) & \jac{g}{y,\del}(p) \\
  \end{bmatrix} \, 
 \end{aligned}
\end{equation}

Applying the Laplace transform to \eqref{eq:ret_singps}:
\begin{equation}\label{eq:ret_dae:lin:lap}
( 
s \bfb{E}(p) - \bfb{A}_0(p) 
- \sum_{\del=1}^{\mu} \bfb{A}_\del(p) 
e^{-s \tau_\del} 
)
\; \mathcal{L} \{ \xys \} = \bfb{E}(p)\xys(0) 
\end{equation}
where $s$ is a complex variable in the $S$-plane.  The associated eigenvalue problem is
\begin{align}
\label{eq:ret_gep:r}
{\bfb P}(s,p) \; \bfb\phi &= \bfg 0_{\nxy,1} \; , \quad \nxy=n+m 
\end{align}
where any value of $s$ that satisfies
\eqref{eq:ret_gep:r}
is an eigenvalue of the system:
\begin{align}
\label{eq:mul_ret_dae:pencil:sparse}
{\bfb P}(s,p) =  s \bfb{E}(p) - \bfb{A}_0(p) - \sum_{\del=1}^{\mu} \bfb{A}_\del(p) e^{-s \tau_\del}  
\end{align}
with $\bfb\phi$ being the associated right eigenvector.  The matrix function ${\bfb P}(s,p)$ plays an important role in the stability and dynamic behavior of 
\eqref{eq:ret_singps}.

\section{Eigenvalue tracking}
\label{sec:eigtrack}

This section presents a continuation-based approach for tracking the eigenvalues of power systems with time delays. The method is first introduced for a single delayed variable, then generalized to systems with multiple delays, and extended to allow the delay magnitude itself to vary as the continuation parameter. 
The formulation is also extended to account for 
realistic time-varying communication delays, using a \acf{wams}
delay model that accounts for stochasticity and data packet dropouts.
Finally, the limitations of the proposed approach are discussed.
\color{black}

\subsection{Single Delay}
\label{sec:tracksingle}

We begin, for simplicity, by considering a single delay $\tau=\tau_1$ ($\mu=1$).  Equation \eqref{eq:mul_ret_dae:pencil:sparse} takes the form:
\begin{equation}\label{eq:sin_ret_dae:pencil:sparse}
{\bfb P} = s \bfb E - \bfb A_0 -
\bfb A_1 e^{-s \tau}
\end{equation}
where the dependence on $p$ is omitted for brevity.  Differentiation of \eqref{eq:ret_gep:r} with respect to $p$ gives:
\begin{equation}
\label{eq:ret_gep:diff1}
\begin{aligned}
\bfb P'
\; \bfb\phi 
+ {\bfb P}
\; \bfb\phi' 
&= \bfg 0_{\nxy,1} 
\end{aligned}
\end{equation}
where 
\begin{equation}
\label{eq:ret_gep:diff2}
\begin{aligned}
\bfb P'
&= s' \bfb E +
{s} \bfb E'
- \bfb A'_0 - \bfb A'_1 e^{-{s}\tau} +\tau s' \bfb A_1 e^{-{s}\tau}
\end{aligned}
\end{equation}
with:
\begin{align*}
s' = \partial s / \partial p \ , \quad
\bfb E' &= \partial {\bfb E} / \partial p\ , \quad \bfb A'_0 = \partial {\bfb A}_0 / \partial p\ , \\ \bfb A'_1 &= \partial {\bfb A}_1 / \partial p\ , \quad
\bfb\phi' = \partial {\bfb\phi} / \partial p \, .    
\end{align*}

%
Equivalently, equation \eqref{eq:ret_gep:diff1} is:
\begin{equation}\label{eq:ret_gep:diff3}
\begin{aligned}
( s \bfb E - \bfb A_0 - \bfb A_1 e^{-s \tau}) \bfb\phi' + (\bfb E + \tau \bfb A_1 e^{-s \tau}) \bfb\phi s'   = 
& \\
- ( s \bfb E' - \bfb A'_0 - \bfb A'_1 e^{-s \tau}) 
\bfb\phi &
\end{aligned}
\end{equation}
System \eqref{eq:ret_gep:diff3}, which describes the evolution of a single eigenpair with respect to $p$, consists of $r$ equations and $r+1$ unknowns, $s$ and $\bfb\phi$. 
To make this system well-posed, it is closed by 
imposing the following eigenvector normalization:
\begin{align}
\label{eq:phi:norm}
\bfg{\phi}\T\bfg{\phi} = 
c
\end{align}

\noindent
where $c$ is an imposed constant, e.g.,~$c=1$. 
Differentiation of \eqref{eq:phi:norm} gives:
%
%
\begin{equation}
\label{eq:phi:norm_dif}
\begin{aligned}
\bfg{\phi}\T \bfg{\phi}' &= 0
\end{aligned}
\end{equation}
Combining \eqref{eq:ret_gep:diff3} and \eqref{eq:phi:norm_dif}:
\begin{align}
\nonumber
\begin{bmatrix}
s \bfb E - \bfb A_0 - \bfb A_1 e^{-s \tau} 
& (\bfb E + \tau \bfb A_1 e^{-s \tau})\bfb\phi \\
{\bfb\phi}\T & 0 
\end{bmatrix}
\begin{bmatrix}
\bfb\phi' \\
 s' 
\end{bmatrix}
= &\\
\begin{bmatrix}
- ( {s} \bfb E' - \bfb A'_0 - \bfb A'_1 e^{-s \tau}) {\bfb\phi} \\
0
\end{bmatrix}
\label{eq:ret_gep:dyn}
\end{align}

Splitting real and imaginary parts, i.e.
\begin{align*}
\bfb\phi &= \bfb\phi_{\rm r} + \jj \bfb\phi_{\rm i} \ , \\ 
s &= s_{\rm r} + \jj s_{\rm i} \ ,
\end{align*}
%
and by setting $\bfb{A}_1 e^{-s_{\rm r} \tau}\cos{(s_{\rm i} \tau)} = \Cs$, $\bfb{A}_1 e^{-s_{\rm r} \tau}\sin{(s_{\rm i} \tau)} = \Sn$, $\bfb{A}'_1e^{-s_{\rm r} \tau}\cos{(s_{\rm i} \tau)} = \Csd$ and $\bfb{A}'_1e^{-s_{\rm r} \tau}\sin{(s_{\rm i} \tau)} = \Snd$ we arrive at the following system:
\begin{equation}
\label{eq:sin_ret_singlevector:sys}
\bfb M(\bfb y) \;  {\bfb y}' = \bfb h (\bfb y)
\end{equation}
where $\bfb y = \bfb y(p) =(\bfb\phi_{\rm r}, \bfb\phi_{\rm i}, s_{\rm r}, s_{\rm i})$, with $\bfb y \in \mathbb{R}^{2\nxy + 2 } $; and:
\begin{align*}
\bfb M(\bfb y) &=
\begin{bmatrix}
    \bfb M_{1} & \bfb M_{2} \\
    \bfb M_{3} & \bfb 0_{2,2}
\end{bmatrix} 
\end{align*}
The quantities $\bfb h$, $\bfb M_{1}$, $\bfb M_{2}$, $\bfb M_{3}$ are defined as:
\begin{equation*}
\begin{aligned}
      \bfb h
    & = 
    \begin{bmatrix}
        (-s_{\rm r}\bfb{E}' + \bfb A'_0 + \Csd)\bfb\phi_{\rm r} + (s_{\rm i}\bfb{E}'+ \Snd)\bfb\phi_{\rm i}\\
        - (s_{\rm i}\bfb{E}'+ \Snd)\bfb\phi_{\rm r} + (-s_{\rm r}\bfb{E}'+ \bfb A'_0 + \Csd)\bfb\phi_{\rm i}\\ 
        0
        \\ 
        0
    \end{bmatrix} \\
\bfb M_{1} &=
    \begin{bmatrix}
        s_{\rm r}\bfb{E}-\bfb{A}_0 - \Cs & -s_{\rm i} \bfb{E} - \Sn \\
        s_{\rm i} \bfb{E} + \Sn  & s_{\rm r}\bfb{E} - \bfb{A}_0 - \Cs
\end{bmatrix}
\\
    \bfb M_{2} &=
    \begin{bmatrix}
      \bfb{E}\bfb\phi_{\rm r}\! +\! \tau( \Cs \bfb\phi_{\rm r}\! +\! \Sn \bfb\phi_{\rm i}) 
      &\!\! -\bfb{E}\bfb\phi_{\rm i}\! -\! \tau( \Cs \bfb\phi_{\rm r}\! -\! \Sn \bfb\phi_{\rm i}) \\
      \bfb{E}\bfb\phi_{\rm i}\! +\! \tau( \Cs \bfb\phi_{\rm r}\! -\! \Sn \bfb\phi_{\rm i}) 
      &\!\! \bfb{E}\bfb\phi_{\rm r}\! +\! \tau( \Cs \bfb\phi_{\rm r}\! +\! \Sn \bfb\phi_{\rm i})
    \end{bmatrix} 
    \\
    \bfb M_{3} &= 
    \begin{bmatrix}
        \bfb\phi_{\rm r}\T & -\bfb\phi_{\rm i}\T \\
        \bfb\phi_{\rm i}\T  & \bfb\phi_{\rm r}\T
    \end{bmatrix}\hspace{49mm}
  \end{aligned}
\end{equation*}

\subsection{Generalization to Multiple Delays}
\label{sec:trackmul}

When multiple delays are present in the system, the expression \eqref{eq:ret_gep:diff1} holds, where in this case the matrix function $\bfb P$ is given by \eqref{eq:mul_ret_dae:pencil:sparse}.
Differentiating \eqref{eq:mul_ret_dae:pencil:sparse} with respect to the continuation parameter $p$ gives:
\begin{equation}
\begin{aligned}
\bfb P'
&= 
s' \bfb E +
{s} \bfb E' - \bfb A'_0 - \sum_{\del=1}^{\mu}
(
\bfb A'_\del - \tau_\del s' \bfb A_\del )e^{-{s}\tau_\del}
\end{aligned}
\end{equation}
Substituting into \eqref{eq:ret_gep:diff1} yields:
\begin{align}
\nonumber
\hspace{-2mm}
(\bfb E + \sum_{\del=1}^{\mu}\tau_\del & \bfb A_\del e^{-s \tau_\del} )  \bfb\phi s' + (
{s} \bfb E - \bfb A_0 - \sum_{\del=1}^{\mu} \bfb A_{\del} e^{-{s} \tau_\del }) \bfb\phi' \\
&= - ( {s} \bfb E' - \bfb A'_0 - \sum_{\del=1}^{\mu}\bfb A'_\del e^{-s \tau_\del}) 
\bfb\phi 
\label{eq:mul_ret_gep:diff3}
\end{align}
Combining with \eqref{eq:phi:norm_dif} and
splitting eigenvalue and eigenvector into real and imaginary parts leads to a system in the form of \eqref{eq:sin_ret_singlevector:sys}, where $\bfb M$ and $\bfb h$ are defined to account for all $\mu$ delays. Their detailed definitions are provided in the Appendix.

\subsection{Delay Magnitude as Varying Parameter}
\label{sec:trackdel}

In this subsection, we track the evolution of eigenvalues as the delay magnitude is varied.  We first consider the case where the system contains a single delayed variable, whose time delay has a varying magnitude and is denoted by $\tau_\ell$.  The corresponding delayed state matrix is denoted by $\bfb A_\ell$. Given $p=\tau_\ell$, the matrix function ${\bfb P}$ is:
\begin{equation}
\label{eq:ddae_tau_pencil}
{\bfb P} = s \bfb E - \bfb A_0 - 
\bfb A_{\ell} e^{-{s} p} 
\end{equation}
%
Matrices $\bfb E$, $\bfb A_0$ and $\bfb A_{\ell}$ do not explicitly depend on $\tau$, hence differentiation of \eqref{eq:ddae_tau_pencil} with respect to $p$ yields:
%
\begin{align}
\nonumber
\bfb P' &= s' \bfb E  
+ \bfb A_\ell(s' p + s) e^{-{s} p}
\end{align}
%
Substituting into \eqref{eq:ret_gep:diff1} yields:
%
%
\begin{equation}
\begin{aligned}
(s \bfb E - \bfb A_0 - \bfb A_{\ell} e^{-s p}) \bfb\phi' 
+ (\bfb E + \bfb A_\ell p e^{-s p}) \bfb\phi s' = \\ - \bfb A_\ell e^{-s p} s  \bfb\phi
\label{eq:sin_ret_gep:tau:der}
\end{aligned}
\end{equation}

By splitting real and imaginary parts of \eqref{eq:sin_ret_gep:tau:der}, we arrive at a system in the form of \eqref{eq:sin_ret_singlevector:sys}, where $\bfb M$ and $\bfb h$ are in this case functions of the varying delay magnitude $p=\tau_\ell$.  Their definitions are detailed in the Appendix.

We extend the above formulation to systems with multiple delays, treating one of them as the continuation parameter, i.e., $\tau_\ell = p$.  The definition of $\bfb P$ in this case is as follows:
\begin{align}
\label{eq:mul_tau_ddae_pencil}
\bfb P = s \bfb E - \bfb A_0 
- \sum_{\del=1}^{\mu} 
\bfb A_{\del} e^{-{s} \tau_\del} 
- \bfb A_{\ell} e^{-{s} p} 
\end{align}
We note that the term corresponding to $\tau_\ell$ is kept separate, even if $\tau_\ell$ equals some $\tau_\del, \;\del=1,2,\ldots,\mu$. Given that 
the matrices $\bfb E$, $\bfb A_0$, $\bfb A_\ell$ and $\bfb A_\del$ do not explicitly depend on $\tau_\ell$, the derivative of \eqref{eq:mul_tau_ddae_pencil} with respect to $p$ is:
%
%
%
\begin{equation}
\bfb P' 
=  s' \bfb E + \sum_{\del=1}^{\mu}\tau_\del s' \bfb A_\del e^{-{s} \tau_\del} + \bfb A_\ell(s' p +s) e^{-{s} p}
\end{equation}
and \eqref{eq:ret_gep:diff1} is equivalently written as:
%
\begin{align}
\nonumber
&(s \bfb E - \bfb A_0 - \bfb A_{\ell} e^{-{s} p } - \sum_{\del=1}^{\mu}\bfb A_{\del} e^{-{s} \tau_\del }) \bfb\phi' +
\\
& (\bfb E + \bfb A_\ell \, p \, e^{-s p} + \sum_{\del=1}^{\mu}\tau_\del \bfb A_\del e^{-s \tau_\del}) \bfb\phi s' = 
- \bfb A_\ell e^{-s p} s \bfb\phi
\label{eq:mul_tau_ret_gep}
\end{align}
By splitting the eigenvalue and eigenvector into real and imaginary parts we arrive at a system in the form of \eqref{eq:sin_ret_singlevector:sys}. 
The derivation of $\bfb M$ and $\bfb h$ is provided in the Appendix.

\subsection{Communication Delays with Noise and Data Dropouts}
\label{sec:wams_del}

The derivations above consider constant delays.  In this section, we show how they can be generalized to capture the effects of realistic \ac{wams} latencies with noise and data packet dropouts.  To this end, we consider the composite time-varying delay model proposed in~\cite{liu2018stability}, in which, for the purpose of \ac{sssa}, packet dropouts and noise are represented through the functions $h_p(s)$ and $h_s(s)$, respectively:
%
\begin{align}\label{eq:hp}
h_p(s) &= \frac{1-p_{dr}}{sT} [1+(p_{dr}-1) \frac{e^{-s T}}{1-p_{dr}e^{-s T}}] \\
\label{eq:hs}
h_s(s) &= (1+\frac{\alpha}{1-p_{dr}} s)^{-b}
\end{align}
where $\alpha$ and $b$ are the scale and shape factors of the \textit{Gamma distribution}; $p_{dr}$ is the packet dropout rate; and $T$ is the normal delivery period for each data packet.

Equation \eqref{eq:mul_ret_dae:pencil:sparse} takes the adjusted form:
\begin{equation}\label{eq:sin_stoc_dae:pencil:sparse}
{\bfb P} = s \bfb E - \bfb A_0 -
h_p(s) h_s(s) \bfb A_1 e^{-s \tau_0}
\end{equation}
where $\tau_0$ is the constant component of the \ac{wams} delay. By setting ${\rm S}_T= h_p(s) h_s(s) e^{-s \tau_0}$, the derivative of \eqref{eq:sin_stoc_dae:pencil:sparse} with respect to $p$ is:
\begin{equation}
\label{eq:stoc_gep:diff2}
\begin{aligned}
\bfb P'
&= s' \bfb E +
{s} \bfb E'
- \bfb A'_0 - {\rm S}'_T  \bfb A_1 - {\rm S}_T  \bfb A'_1
\end{aligned}
\end{equation}
where 
%
\begin{equation}
\label{eq:dot_S_T_full}
\begin{aligned}
{\rm S}'_T = 
 & (\frac{\partial{h_p}}{\partial{p}} h_s + h_p \frac{\partial{h_s}}{\partial{p}}) e^{-s \tau_0} + \\
& (\frac{\partial{h_p}}{\partial{s}} h_s + h_p \frac{\partial{h_s}}{\partial{s}} 
- \tau_0 h_p h_s ) e^{-s \tau_0} s'\\
= & {\rm S}_{Tp} + {\rm S}_{TD}s'
\end{aligned}
\end{equation}
%
\color{black}
We note that for a constant delay, i.e., $\tau(t)=\tau_0$, the model reduces to $h_p(s)=h_s(s)=1$ and $\partial{h_p}/\partial{s}=\partial{h_s}/\partial{s}=0$; therefore, equations \eqref{eq:sin_stoc_dae:pencil:sparse} and \eqref{eq:stoc_gep:diff2} are simplified to \eqref{eq:sin_ret_dae:pencil:sparse} and \eqref{eq:ret_gep:diff2} respectively.

Substituting \eqref{eq:sin_stoc_dae:pencil:sparse} and \eqref{eq:stoc_gep:diff2} into \eqref{eq:ret_gep:diff1} yields:
%
\begin{equation}\label{eq:wams_diff}
\begin{aligned}
( s \bfb E - \bfb A_0 - {\rm S}_T \bfb A_1) \bfb\phi' + (\bfb E - {\rm S}_{TD} \bfb A_1) \bfb\phi s'  &  = \\ 
- (s \bfb E' - \bfb A'_0 - \bfb{S}_{Tp} \bfb A_1 - \bfb{S}_{T} \bfb A'_1) \bfb\phi &
\end{aligned}
\end{equation}
By splitting real and imaginary parts of \eqref{eq:wams_diff}, we arrive at a system in the form of \eqref{eq:sin_ret_singlevector:sys}, where the definitions of $\bfb M$ and $\bfb h$ are detailed in the Appendix.


Table~\ref{tab:comparison} summarizes the key differences among the proposed approach and the ones proposed in \cite{mou2021delayed} and \cite{li2017eigenvalue}.
%



\begin{table}[htbp]
\centering
\caption{
Comparison of eigenvalue tracking approaches.}
\label{tab:comparison}
\small
\setlength{\tabcolsep}{4pt}
\renewcommand{\tabularxcolumn}[1]{m{#1}}
\begin{tabularx}{\linewidth}{
>{\raggedright\arraybackslash}X
>{\centering\arraybackslash}X
>{\centering\arraybackslash}X
>{\centering\arraybackslash}X}
\toprule
\textbf{Feature} &
\textbf{\cite{mou2021delayed}} &
\textbf{\cite{li2017eigenvalue}} &
\textbf{Proposed} \\
\midrule

Delay model &
constant &
constant &
constant, time-varying, stochastic \\
\midrule
Continuation parameter &
system/control parameters, delays &
delays &
system/control parameters, loads, delays, non-constant delay parameters \\
\midrule
Steady-state solution &
assumed fixed &
assumed fixed &
updates the solution on every step
\\
\midrule
Predictor &
1st-order sensitivity &
previous eigenpair &
numerical integration \\
\midrule
Corrector (local convergence) &
quasi-Newton (linear) &
Newton (quadratic) &
Newton (quadratic)\\
\bottomrule
\end{tabularx}
\end{table}


\color{black}

\subsection{Numerical Integration}

Given a parameter range $[p_{init},p_{fin}]$, tracking is performed by numerically integrating the parameter-space system \eqref{eq:sin_ret_singlevector:sys}. For the initial value $p_{init}$, the state vector $\bfb y(p_{init})$ is obtained by solving the eigenvalue problem \eqref{eq:ret_gep:r}.  An effective method, in our experience, is to transform the linearized \acp{ddae} into an equivalent system of partial differential equations, and then reduce it through Chebyshev polynomials to a finite-dimensional linear eigenvalue problem.  The spectral discretization technique used in this paper is described in \cite{Li_discr}.  
In subsequent steps, 
matrix derivatives are computed numerically using first-order finite differences.
If the varying parameters alter the power flow solution the latter is recomputed.
For the case of the \ac{aiits} described in Section~\ref{sec:case_study}, 
Newton's method typically converges within 3-5 iterations and it requires less than $0.003$~s in each continuation step, increasing the total computational time by less than $1\%$.  Thus, in practice, the power flow is recomputed at every step without a noticeable impact on the method's speed.
Once the operating point has been updated the matrices $\bfb A_0$, $\bfb A_1$ and $\bfb E$ are reconstructed.
\color{black}

Integration proceeds iteratively until $p=p_{fin}$. First, the parameter is updated as $p_{k+1} = p_{k} + \Delta p$, where $\Delta p$ is the step size.
The mass matrix $\bfb M(\bfb{y}_k) $ and right-hand side $ \bfb h(\bfb{y}_k)$ are then evaluated, and the state vector $\bfb{y}$ is advanced using a numerical integration method.  For instance, using the \ac{fem} gives:
\begin{equation}  \label{eq:singlevector:fem}
  \bfb{y}_{k+1} = \bfb{y}_k + \Delta p \,\bfb M^{-1}(\bfb{y}_k) \, \bfb h (\bfb{y}_k)
\end{equation}
The product $\bfb M^{-1}(\bfb{y}_k) \,\bfb h (\bfb{y}_k)$ can be determined through LU decomposition of $\bfb M(\bfb{y}_k)$.

The choice of integration method for \eqref{eq:sin_ret_singlevector:sys} involves a trade-off between speed and accuracy; 
for example, 
the \ac{fem} is conditionally stable and its local truncation error decreases quadratically with the step size, for sufficiently small continuation steps. The \ac{tm} is symmetrically A-stable, and its local truncation error decreases cubically with the step size, for sufficiently small continuation steps. It generally provides higher accuracy than \ac{fem}, at the expense of additional computational cost \cite{TZOUNAS2022108266}.
\color{black}
We note that regardless of the choice, numerical integration schemes inherently introduce discretization error, which depends on the step size.  
This error can be effectively mitigated by adding a Newton-based corrector step that refines the solution to a prescribed tolerance, at additional computational cost. 
The latter depends on the number of iterations required to achieve the prescribed tolerance, as each iteration requires a sparse LU factorization.
Combining equations \eqref{eq:ret_gep:r} and \eqref{eq:phi:norm} for $c=1$, the Newton-based corrector step is formulated as follows:
\begin{equation}
\begin{bmatrix}
\frac{\partial \bfg f}{\partial \bfb\phi}
& \frac{\partial \bfg f}{\partial s}\\
\hat{\bfb\phi}\T & 0 
\end{bmatrix}
\begin{bmatrix}
\Delta \bfb\phi \\
 \Delta s 
\end{bmatrix}
= 
- \begin{bmatrix}
\bfg f(s^{[j]}, \bfb\phi^{[j]}) \\
\hat{\bfb\phi}\T \bfb\phi^{[j]} -1
\end{bmatrix}
\label{eq:NR}
\end{equation}
where $\bfg f = \bfb P \cdot \bfb\phi$ and $j$ is a positive integer representing the $j^{\rm th}$ iteration. The Newton corrector terminates when the norms $||\bfg f(s^{[j]}, \bfb\phi^{[j]})||$ and $||\hat{\bfb\phi}\T \bfb\phi^{[j]} -1||$ fall below the prescribed tolerance, or when a prescribed maximum number of iterations is reached.
For the results presented in Section~\ref{sec:case_study}, 
a Newton corrector with an error tolerance of $10^{-8}$ is applied after each step, typically requiring 2-5 iterations to converge.
If the maximum of $10$ iterations is reached, the corrector stops, $\Delta p$ is halved, and the last predictor step is repeated.
\color{black}

\subsection{Discussion on  Practical Limitations}

While the proposed approach offers significant advantages in terms of computational efficiency and robustness, its practical application has certain limitations.
The main limitation of the current formulation is that it is restricted to tracking a single eigenpair at a time; thus studying multiple modes simultaneously requires an equal number of independent runs. Although these can be performed in parallel, extending the formulation to simultaneously track multiple modes remains a topic for future work.
%

Other potential challenges relate to the computational cost of the proposed approach. The latter can be increased in the case that sparse matrix factorizations lead to a substantial fill-in. 
However, this is generally limited in power system models due to their highly sparse structure.
%
%
Another case posing numerical challenges is when eigenvalues coalesce. As two eigenvalues approach one another, their sensitivity to the continuation parameter grows large. This can cause the predictor to supply the Newton corrector with a poor initial guess and require additional iterations to converge. Moreover the continuation equations may in theory become ill-conditioned in the vicinity of a coalescence point, where the eigenvalue sensitivity is not defined. The above challenges are accounted for in the implementation of the proposed approach.
%
\color{black}

\section{Case Study}
\label{sec:case_study}

In this section, we apply the continuation-based eigenvalue tracking formulation of Section~\ref{sec:eigtrack} to two power system models.  
First, we consider a modified version of the IEEE 39-bus system based on the one reported in~\cite{hiskens2013ieee}, 
\color{black}
and then a real-world-scale model of the Irish transmission system. 
All simulations are carried out with the power system software tool Dome \cite{dome}.  The version of the software employed in this paper relies on LAPACK~3.12.0 for QR factorization
and KLU~2.3.6 for LU factorization. For the sparse \ac{ks} method it relies on SLEPc~3.24.1~\cite{slepc}.
\color{black}

\subsection{39-Bus System}

We consider a modified version of the IEEE 39-bus system, wherein the machines at buses 30, 34, 35, 36, 37 are replaced by \acfp{der} of equal capacity. 
For the purposes of this study, we reduce the inertia constant of the \ac{sg} connected to bus~39 by a factor of ten. Apart from that, for each synchronous generator in the modified system, the parameters corresponding to the same generator in the benchmark system are retained.
Each \ac{der} comprises an inner control loop that regulates the $d$ and $q$ components of the current in the $dq$ reference frame, and two outer loops for primary frequency and voltage control, respectively
\cite{ORTEGA201837}.
The \ac{der} controller parameters are listed in Table~\ref{tab:mod39b_param}. $\rm T_d$ and $\rm T_q$ are the $d$- and $q$- axis current-controller time constants, respectively, and $\rm R$, $\rm T_f$, and $\rm T_w$ denote the droop gain, droop time constant, and washout filter time constant, respectively.
A \ac{pss} is also added to each of the five \acp{sg}, retaining the control parameters given in \cite{hiskens2013ieee}. Each \ac{pss} receives the local frequency deviation as input and modulates the field excitation voltage of the corresponding \ac{sg} through the reference voltage of its \ac{avr}.

\begin{table}[!ht]
\caption{
Parameters of DER controllers.}
\label{tab:mod39b_param}
\centering
\begin{tabular}{c|lll}
\toprule
\centering
 {Controller}  & \multicolumn{3}{c}{Parameters}\\
\midrule
{Current} & $\rm T_d=0.0001$~s & $\rm T_q=0.0001$~s &  \\
{Frequency} & $\rm R=0.05$ & $\rm T_f=1.1$~s & $\rm T_w=0.1$~s\\
{Voltage} & $\rm K_P=10$ & $\rm K_I=5$ &  \\
\bottomrule
\end{tabular}
\end{table}
\color{black}

\begin{figure}[pos=ht]
  \centering 
  \begin{subfigure}{\columnwidth}
    \centering 
  \includegraphics[width=\linewidth]{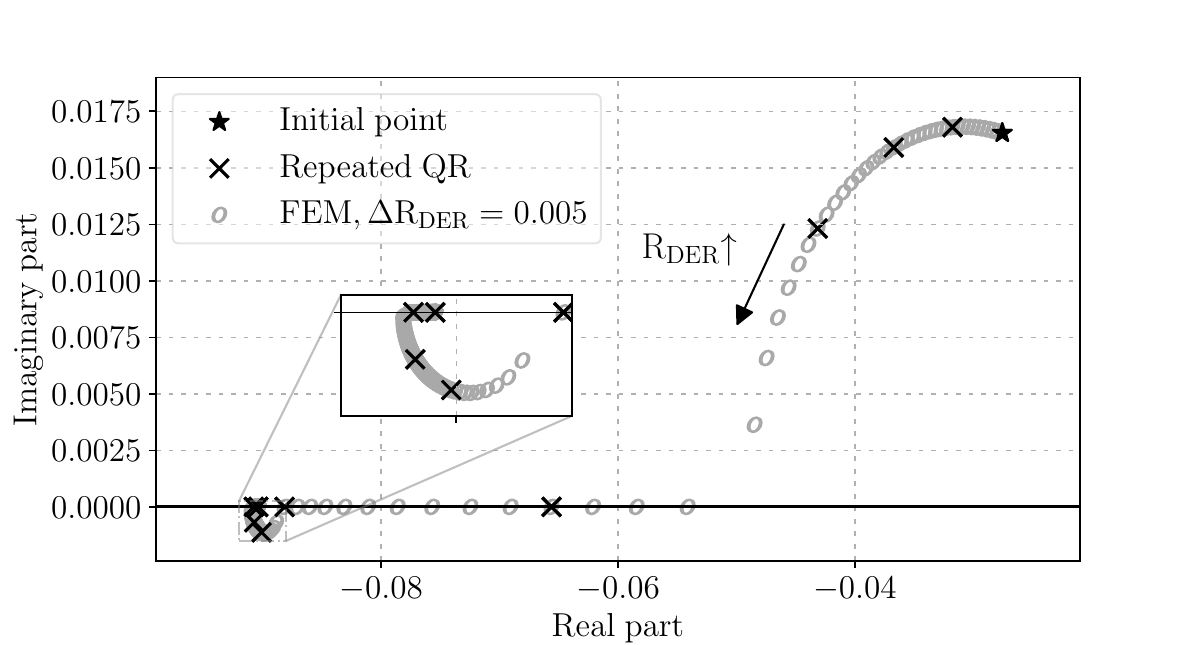}
  \caption{FR mode: increasing the DER frequency controllers' droop constant.}
  \label{fig:mod39b_DER_R_1D_65}
  \end{subfigure}
  \begin{subfigure}{\columnwidth}
    \centering 
  \includegraphics[width=\linewidth]{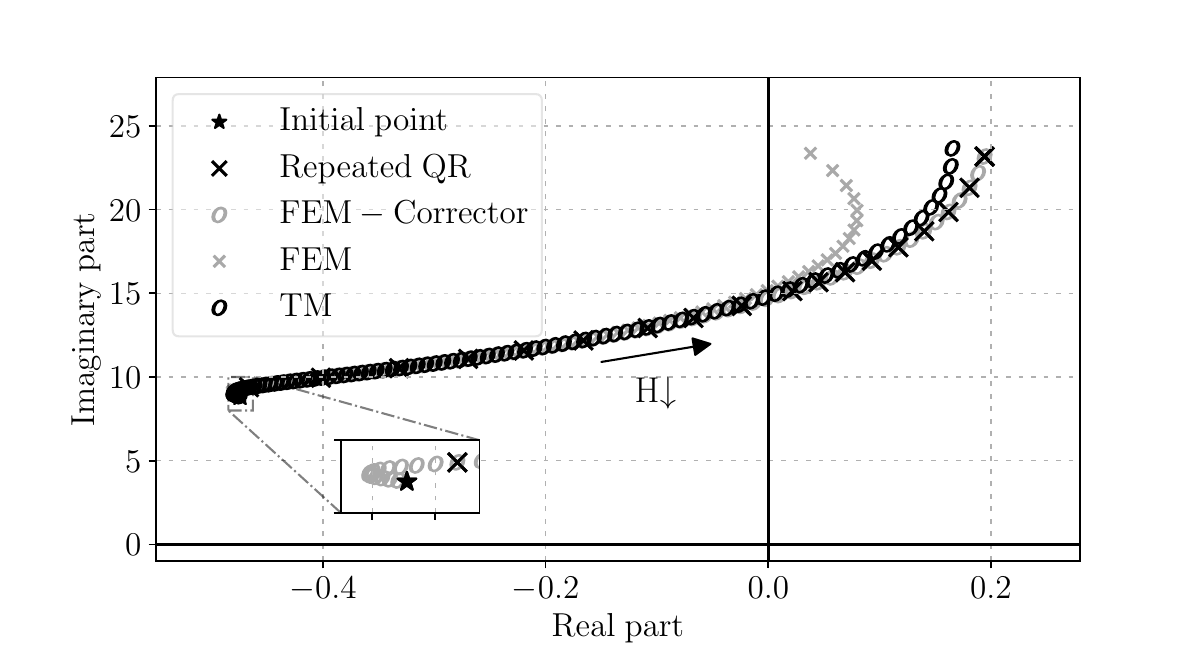}
  \caption{Critical electromechanical mode: Reducing the inertia of the SG at bus~39.}
  \label{fig:mod39b_SYN_H_1D_21}
  \end{subfigure}
  \caption{39-bus system: tracking in the presence of a single delay.}
  \label{mod39b:1D}
\end{figure}

We begin by demonstrating the proposed tracking method on the system's \ac{fr} mode, also referred to as common low-frequency mode in the literature \cite{Bernal_2016, Moeini_2016}.  This is a system-wide mode originating from the dynamics of frequency control loops \cite{bouterakos2025eigenvalue}.
For the examined system, the natural frequency of the \ac{fr} mode is initially $0.005$~Hz.  To illustrate the process, the frequency control droop constant $\rm R_{DER}$ of the five \acp{der} is varied from $0.05$ to $0.5$.  A single delay of $\tau = 0.01$~s is introduced in the frequency signal of the \ac{pss} connected to the \ac{sg} at bus~31.  The small delay is not intended to significantly alter the system dynamics, but to assess the tracking formulation while another system parameter is varied.  The traced eigenvalue, initially at $-0.0276 +\jj 0.0165$, follows the trajectory shown in Fig.~\ref{fig:mod39b_DER_R_1D_65}.  Tracking based on \eqref{eq:sin_ret_singlevector:sys} accurately captures the trajectory and shows how the oscillatory behavior of the \ac{fr} mode is progressively suppressed as the relative share of frequency control shifts from the \acp{der} to the \acp{sg}.  
At the point where the complex pair coalesces on the real axis, the eigenvalue becomes defective with algebraic multiplicity~$2$ and geometric multiplicity~$1$.  This corresponds to a simple quadratic fold \cite{sun1990multiple, bouterakos2025eigenvalue}.  As $\rm R_{DER}$ is further increased, the defective eigenvalue splits into two distinct real eigenvalues.  The tracking method inherently follows one of the two emerging branches, as determined by the evolution of \eqref{eq:sin_ret_singlevector:sys}.  In this example, the eigenvalue on the left branch of the fold is traced, and Fig.~\ref{fig:mod39b_DER_R_1D_65} shows two additional folds occurring along this trajectory.  In practice, when such a splitting occurs, the process is reinitialized by recomputing the eigendecomposition of $\bfb P$ and selecting the eigenpair corresponding to the branch of interest.

\color{black}
As a second example, the inertia constant of the system's largest \ac{sg} is gradually reduced from $50$ to $5$.  Figure~\ref{fig:mod39b_SYN_H_1D_21} illustrates the resulting trajectory of the system's least damped mode, initially located at $-0.4745 \pm \jj 8.8572$ and corresponding to the local electromechanical mode of the \ac{sg} at bus~31.  Tracking based on \eqref{eq:sin_ret_singlevector:sys} shows how the eigenvalue moves as inertia decreases, eventually crossing into the unstable region.
In the same figure we verify that for the same step of $\Delta \rm{H}=-0.5$, the use of \ac{tm} for numerical integration outperforms \ac{fem} in terms of accuracy. Nevertheless, the combined use of the numerical integration step with a Newton corrector tracks the trajectory with the prescribed tolerance.
\color{black}

\begin{figure}[pos=ht]
  \centering 
  \begin{subfigure}[t]{\columnwidth}
    \centering 
  \includegraphics[width=\linewidth]{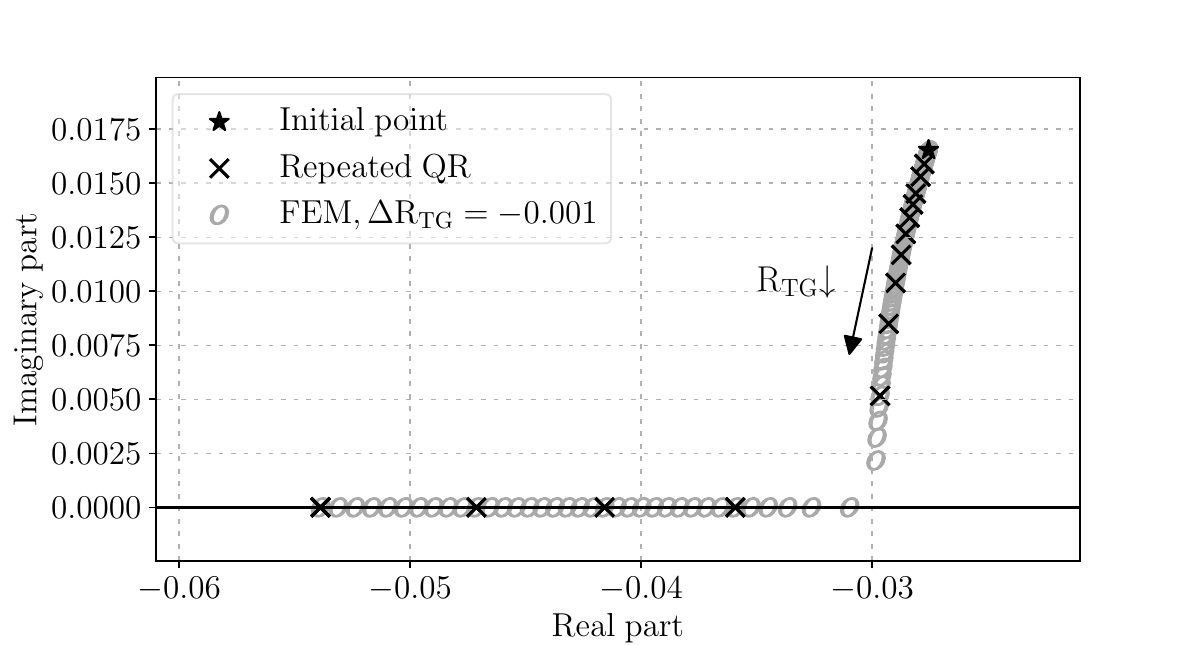}
  \caption{FR mode.}
  \label{fig:mod39b_TG_R_20D_46}
  \end{subfigure}
  \centering
  \begin{subfigure}[t]{\columnwidth}
    \centering 
  \includegraphics[width=\linewidth]{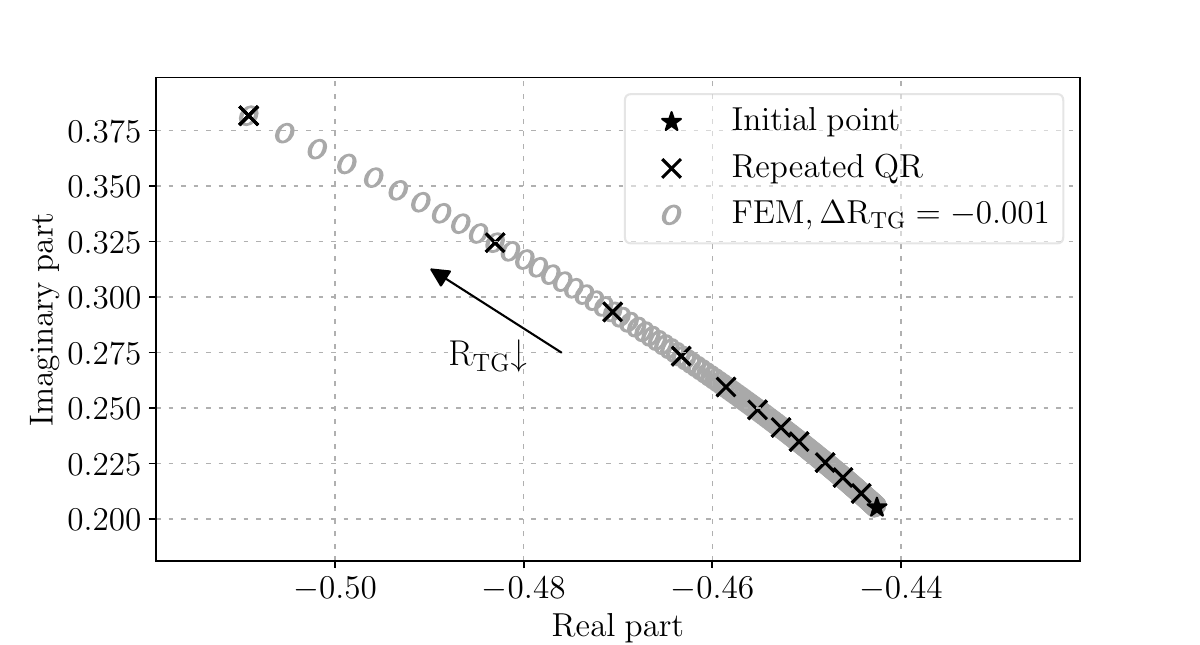}
  \caption{Inter-machine oscillation mode.}
  \label{fig:mod39b_TG_R_20D_39}
  \end{subfigure}
\caption{39-bus system: tracking in the presence of multiple delays.}
\label{mod39b:20D}
\end{figure}

We next assess the method in the presence of multiple delays. To this end, we introduce delays to the input signals of the system's \acp{avr}, \acp{pss} and \ac{der} frequency and voltage controllers, yielding a total of 20 time-delayed variables.  The resulting $\bfb P$ has dimension 558$\times$558.  For each delay $\tau_\del$, the spectral discretization in \cite{Li_discr} uses $N$ Chebyshev nodes in the interval $[-\tau_\del, 0]$, which increases the size of the approximated linear eigenvalue problem by $N+1$.  Thus, considering $N = 10$, the dimension of the approximated linear eigenvalue problem grows by $220$ compared to the delay-free model.  We then trace how variations in the \ac{sg} droop constants affect the eigenvalues corresponding to 
(i) the system's \ac{fr} mode and 
(ii) the inter-machine mode with the largest participation in the rotor speeds of the \acp{sg} at buses $32$ and $39$, initially located at $-0.4425\pm\jj 0.2052$.   The calculated trajectories, shown in Fig.~\ref{mod39b:20D}, confirm that the proposed tracking method accurately captures eigenvalue trajectories in the presence of multiple delays.

\begin{figure}[pos=ht]
  \centering \includegraphics[width=\linewidth]{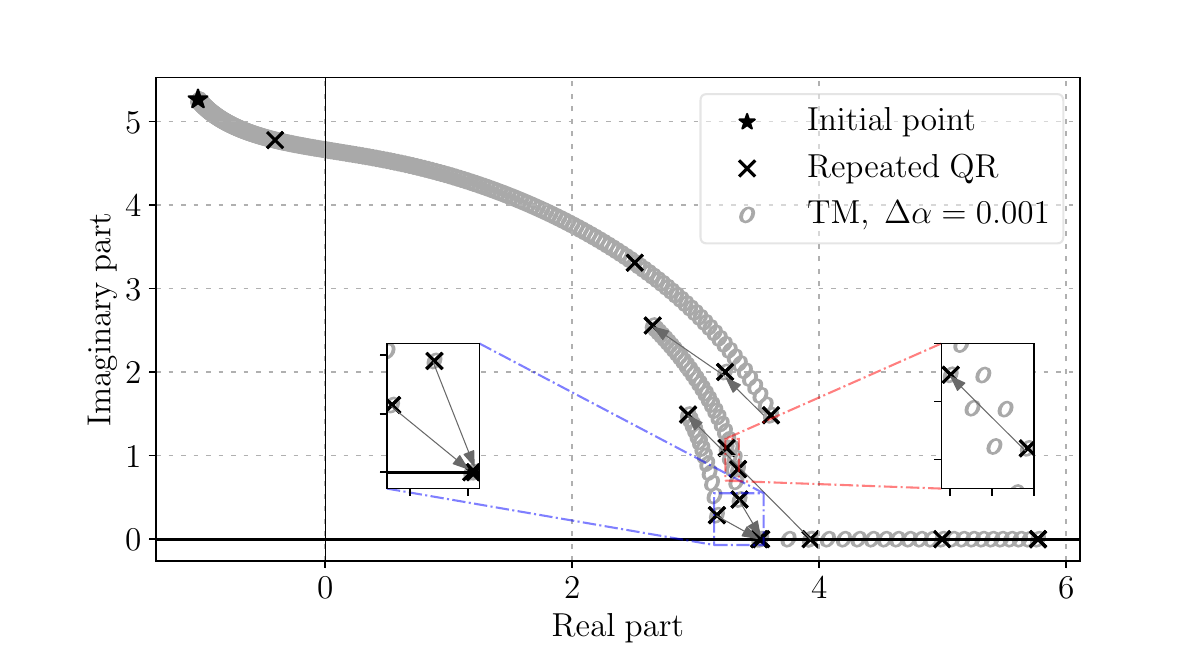}
  \caption{
  39-bus system: effect of increasing the system's loading.
  }  \label{fig:mod39b_IA_loading}
\end{figure}

Next, to assess the accuracy of the method when the system's operating point is altered, we use a loading parameter $\alpha$,
and starting from $\alpha=0$, the active and reactive power of the loads are increased by a factor of $1+\alpha$, up to $\alpha=0.3$,
similar to a continuation power flow process \cite{ajjarapu2002continuation}.   
By solving the power flow at each step, the operating point is updated and the matrices
$\bfb E$ and $\bfb A$ are subsequently reconstructed.
In this case, the input signals of the 5 \acp{pss} are impacted by a constant delay $\tau_{\rm PSS}=0.05$~s and the system's interarea mode is being traced, as it is the most critical one for stability.
Figure~\ref{fig:mod39b_IA_loading} shows the tracking results as the system moves toward its maximum loadability limit.
We observe that as the increased loading causes system components to reach their active or reactive power limits, the mode follows a highly nonlinear trajectory with multiple jumps.
The latter are successfully captured by the proposed method as the operating point changes with increasing system loading.
%

The selection of the method's step size represents a trade-off between accuracy, convergence, and computational time. Smaller step sizes $\Delta p$ reduce the discretization error and generally provide a better initial estimate for the Newton corrector, but increase the number of continuation steps. On the other hand, larger steps reduce the total computational time, but increase the predictor error and may require more Newton iterations or affect convergence.
In practice, this trade-off is strongly influenced by the local eigenvalue sensitivity with respect to the continuation parameter, i.e., $s' = \partial s / \partial p$, which is available directly as a byproduct of the proposed continuation formulation. If $|\Delta s_{\rm tar}|$ is a prescribed target eigenvalue displacement, the step size can be set to $|\Delta s_{\rm tar}| / |s'|$. However, the sensitivity $s'$ can vary significantly along the continuation path.

\begin{figure}[pos=ht]
  \centering \includegraphics[width=\linewidth]{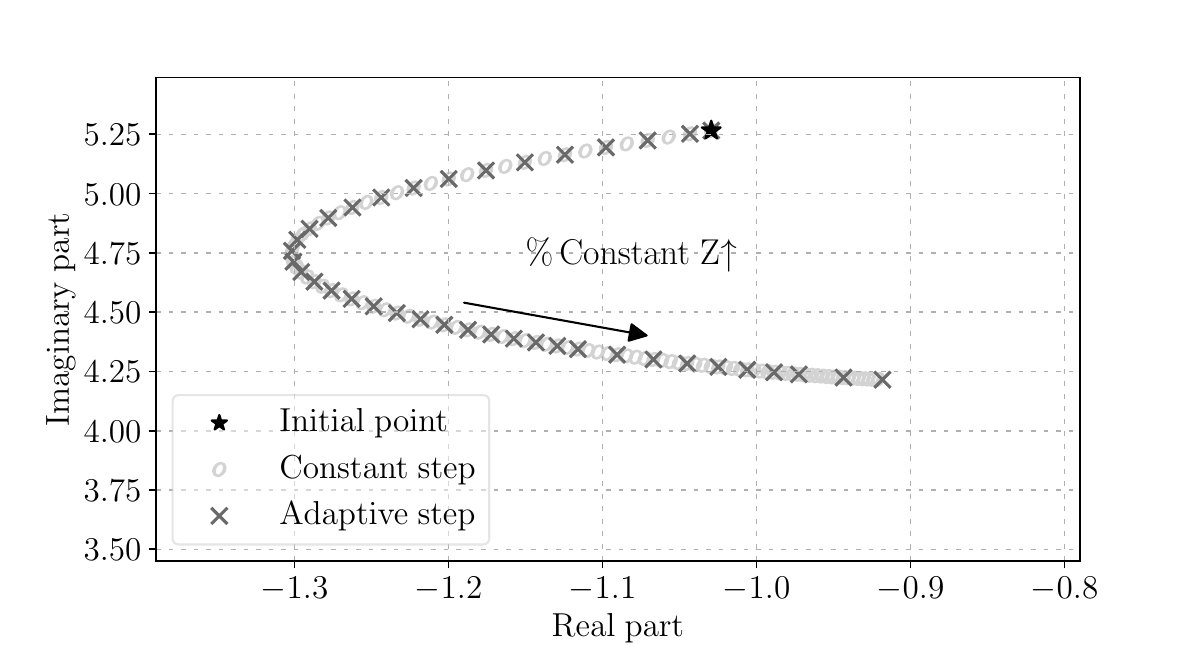}
  \caption{
  39-bus system: effect of varying the share of constant impedance loads of the system from $0$ to $100\%$.
  \color{black}
  }  \label{fig:mod39b_IA_loads}
\end{figure}

In this vein, the efficiency of the proposed method can be further improved by using an adaptive step size strategy. For the purposes of this paper, we consider a heuristic strategy in which the integration step is doubled or halved based on the displacement $|\Delta s|$ between two consecutive continuation steps. 
We introduce a constant delay $\tau_{\rm PSS}=0.05$~s in the input signals of the system's five \acp{pss} and track the interarea mode, initially located at $-1.0294\pm\jj 5.2660$, as the system's constant power loads are gradually replaced by constant impedance (constant Z) ones.
%
The result in Fig.~\ref{fig:mod39b_IA_loads}, indicates that replacing the loads initially improves the damping of the interarea mode, 
but a turning point is eventually reached beyond which the damping decreases.
The integration step $\Delta p$, corresponding to the change in the fraction of constant-power loads, is doubled if $|\Delta s| < 0.02$ and halved if $|\Delta s| > 0.05$.
The adaptive step-size strategy reduces the number of steps by $62\%$. Since the method's computational burden depends on the total number of steps, adopting such strategies can further increase its efficiency. 

\begin{figure}[pos=ht]
  \centering 
  \begin{subfigure}{\columnwidth}
    \centering
    \includegraphics[width=\linewidth]{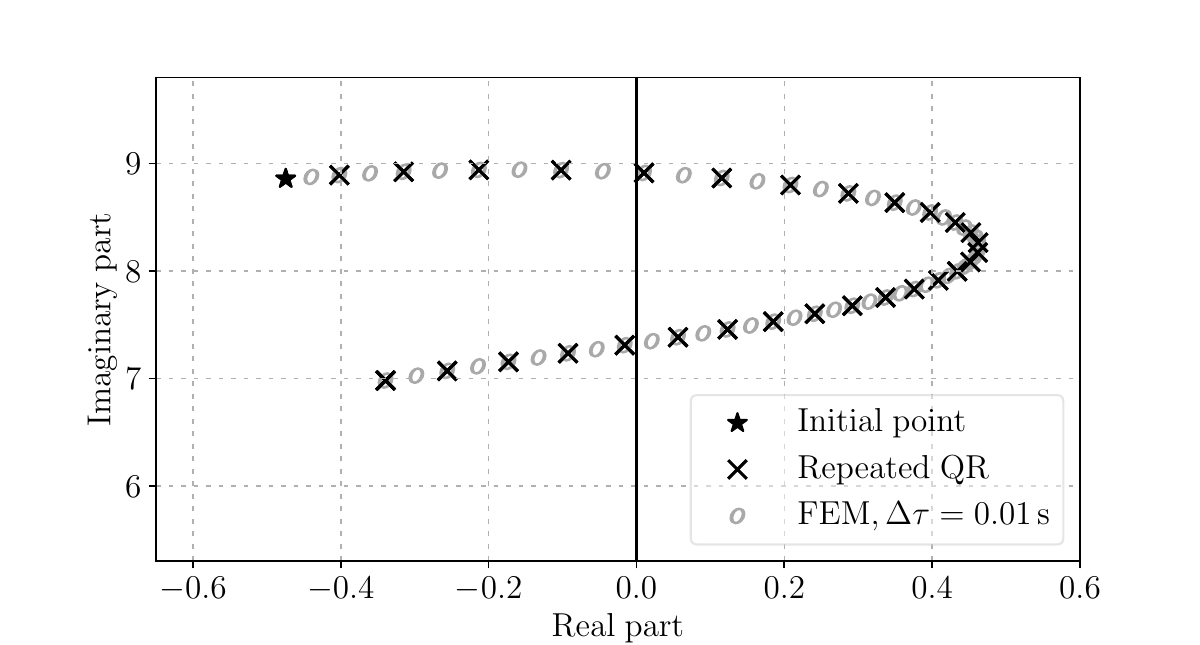}
    \caption{Tracing a local electromechanical mode as delay is increased.}
    \label{fig:mod39b_PSS2_tau}
  \end{subfigure}
  \hfill
  \begin{subfigure}{\columnwidth}
    \centering    \includegraphics[width=\linewidth]{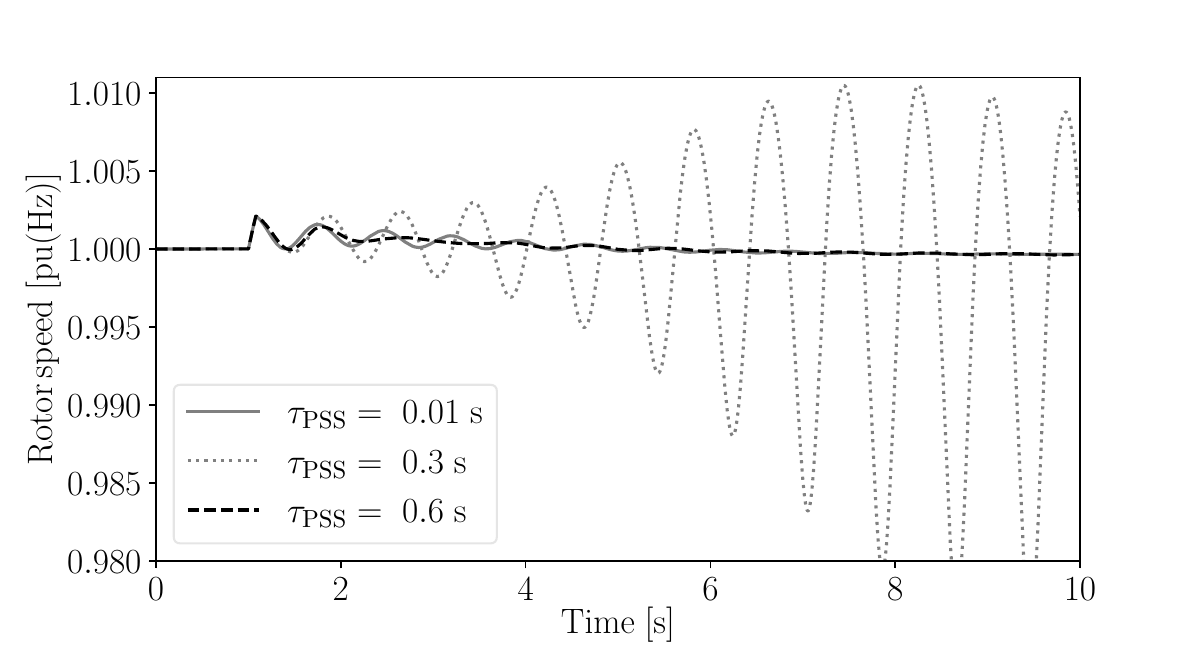}
    \caption{Time-domain simulation for different delay values.}
\label{fig:mod39b_PSS2_tau_TDS}
  \end{subfigure}
  \caption{39-bus system, SG at bus~31: PSS with input signal delay.}
  \label{fig:mod39b_stab_marg}
\end{figure}

\color{black}
We next examine the tracking accuracy when the delay magnitude is treated as the continuation parameter, as described in Section~\ref{sec:trackdel}.  The analysis focuses on the local electromechanical mode of the \ac{sg} at bus~31.  The delay $\tau_{\rm PSS}$ in the input signal of the 
corresponding \ac{pss} is varied from $0.01$ to $0.6$~s, producing the trajectory shown in Fig.~\ref{fig:mod39b_PSS2_tau}.
As $\tau_{\rm PSS}$ increases, the traced eigenvalue becomes progressively less damped and eventually crosses into instability for $\tau_{\rm PSS} > 0.1$~s.  
Further increasing $\tau_{\rm PSS}$ beyond $0.28$~s reverses this trend, and the system eventually regains small-signal stability for $\tau_{\rm PSS} > 0.5$~s.  This behavior is further validated via a time-domain simulation, 
considering a three-phase fault at bus~6, cleared after 80~ms by opening line 5–6. 
The corresponding \ac{sg} rotor-speed response is shown in Fig.~\ref{fig:mod39b_PSS2_tau_TDS} and confirms the conclusions drawn by the eigenvalue tracking analysis.
In the case of $\tau_{\rm PSS}=0.3$~s, the \ac{pss}'s output introduces a destabilizing negative damping component in the \ac{sg}'s electrical torque, causing the rotor speed to diverge.
%
%
Figure~\ref{fig:mod39b_stab_marg} suggests that, under certain conditions, an appropriately selected delay can improve the damping of electromechanical modes. This idea is discussed in detail in \cite{tzounas2021damping}, where it is exploited in the design of a proportional-retarded controller.

\begin{figure}[pos=ht]
  \centering \includegraphics[width=\linewidth]{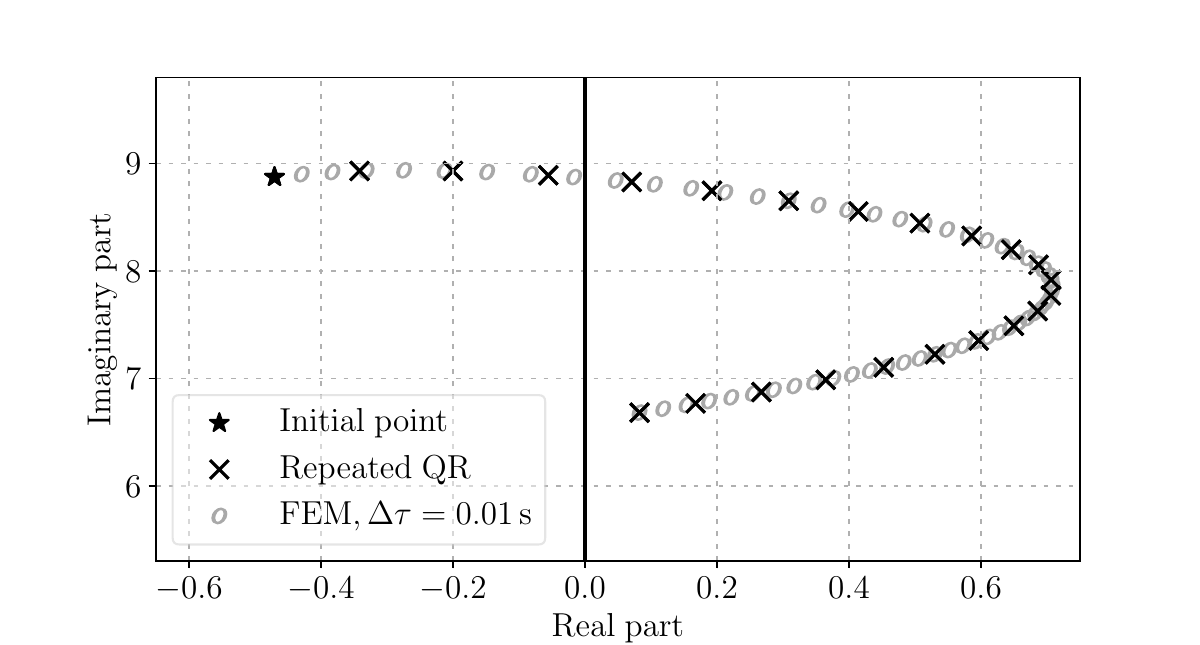}
  \caption{
  39-bus system: effect of increasing the input signal delay of all 5 \acp{pss}.
  \color{black}
  }  \label{fig:mod39b_stab_marg_mult}
\end{figure}

We repeat the tracking for the same mode, when a constant delay is introduced to the input signal of all of the system's five \acp{pss}. The magnitude of the delay is increased in the same range as in Fig.~\ref{fig:mod39b_stab_marg}. The result, shown in Fig.~\ref{fig:mod39b_stab_marg_mult}, depicts a similar trajectory. 
However, the traced eigenvalue crosses into instability for a smaller value of $\tau_{\rm PSS} \approx 0.093$~s. 
We observe that simultaneously delaying the input signals of all five \acp{pss} has a more pronounced effect on the traced mode than delaying a single \ac{pss} signal.
\color{black}

\color{black}

To demonstrate the applicability of the proposed tracking method to \ac{wams} delays, we consider a remote signal used for damping the system's interarea oscillation.
As input to the wide-area \ac{pss}, we select the signal with the highest observability of the interarea mode, namely $\rm \omega_{38-32}=\omega_{38}-\omega_{32}$, i.e., the difference between the rotor speeds of the \ac{sg} at bus $38$ and the one at bus $32$. The \ac{pss} output is added to the reference voltage of the \ac{avr} at bus $38$, which provides the highest controllability of the targeted mode.

\begin{figure}[pos=ht]
  \centering 
  \begin{subfigure}{\columnwidth}
    \centering
    \includegraphics[width=\linewidth]{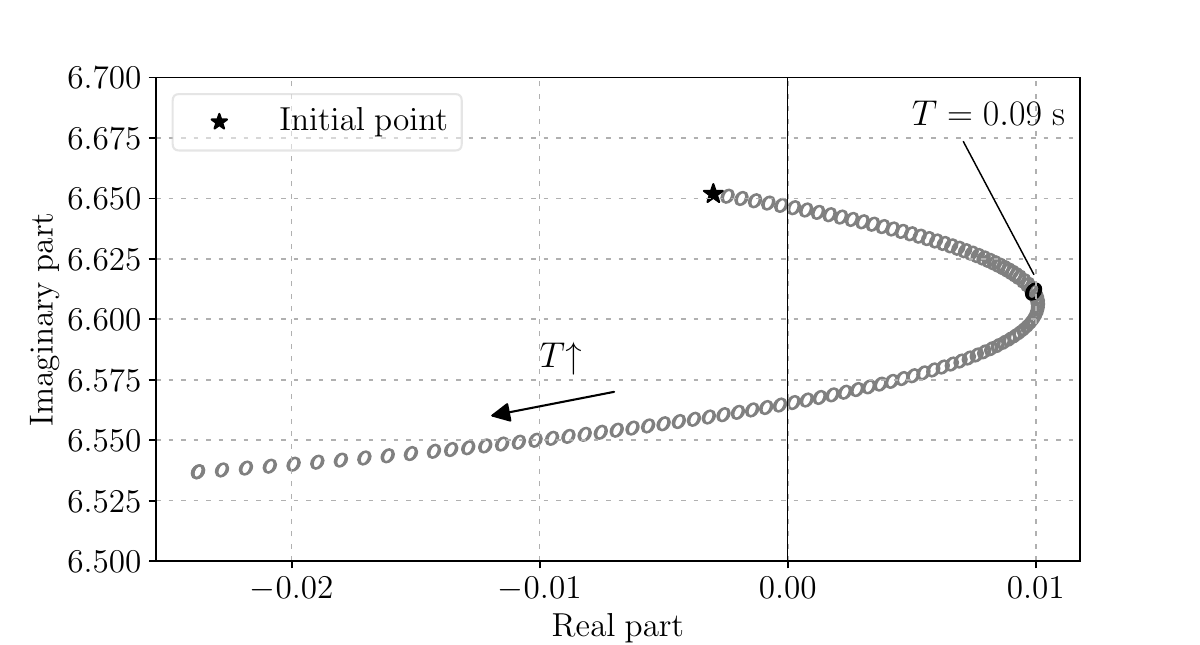}
    \caption{
    Tracing the interarea mode as $T$ is increased.}
    \label{fig:mod39b_wam_Td_track}
  \end{subfigure}
  \hfill
  \begin{subfigure}{\columnwidth}
    \centering    \includegraphics[width=\linewidth]{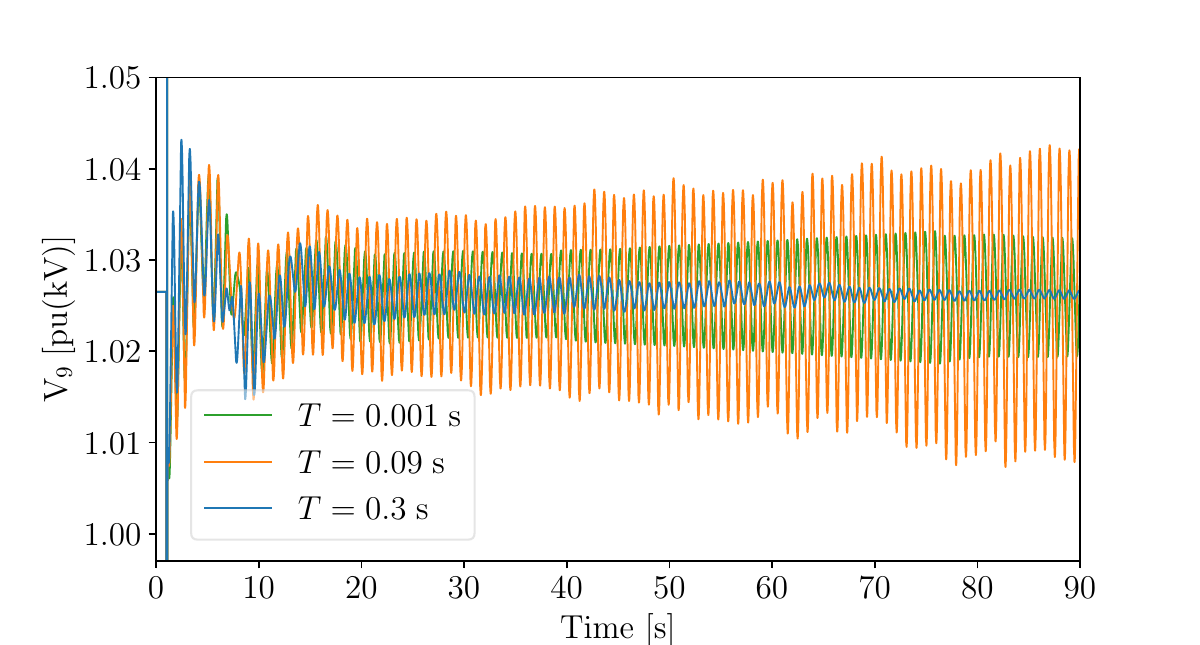}
    \caption{
    Time-domain simulation for different $T$ values.}
\label{fig:wams_tds_Td}
  \end{subfigure}
  \caption{
    39-bus system: effect of the \ac{wams} delay delivery period $T$.}
  \label{fig:mod39b_wam_Td}
\end{figure}

The rotor speed signal $\omega_{32}$ is subject to the latency introduced by the \ac{wams}. We model the \ac{wams} latency using  $\tau_0=0.07$~s, $\alpha=0.01$, and $b=1$, and study the effect of varying the delivery period $T$ and the data packet dropout rate $p_{dr}$. Keeping $p_{dr}$ constant and equal to $0.1$, we increase the normal delivery period from $0.001$ to $0.3$~s while tracing the system's interarea mode, initially located at $-0.00300 \pm\jj 6.65182$. The result is shown in Fig.~\ref{fig:mod39b_wam_Td_track}. The initially stable mode enters the unstable region as it moves rightwards in the complex plane. For $T \approx 0.09$~s, this trend is reversed and the mode moves leftwards, eventually regaining stability. The time-domain simulation shown in Fig.~\ref{fig:wams_tds_Td} considers the same contingency as Fig.~\ref{fig:mod39b_PSS2_tau_TDS}. Considering three distinct values of $T$, i.e., $0.001$, $0.09$ and $0.3$~s, the time-domain simulation validates the tracking results.
%

\begin{figure}[pos=ht]
  \centering 
  \begin{subfigure}{\columnwidth}
    \centering
    \includegraphics[width=\linewidth]{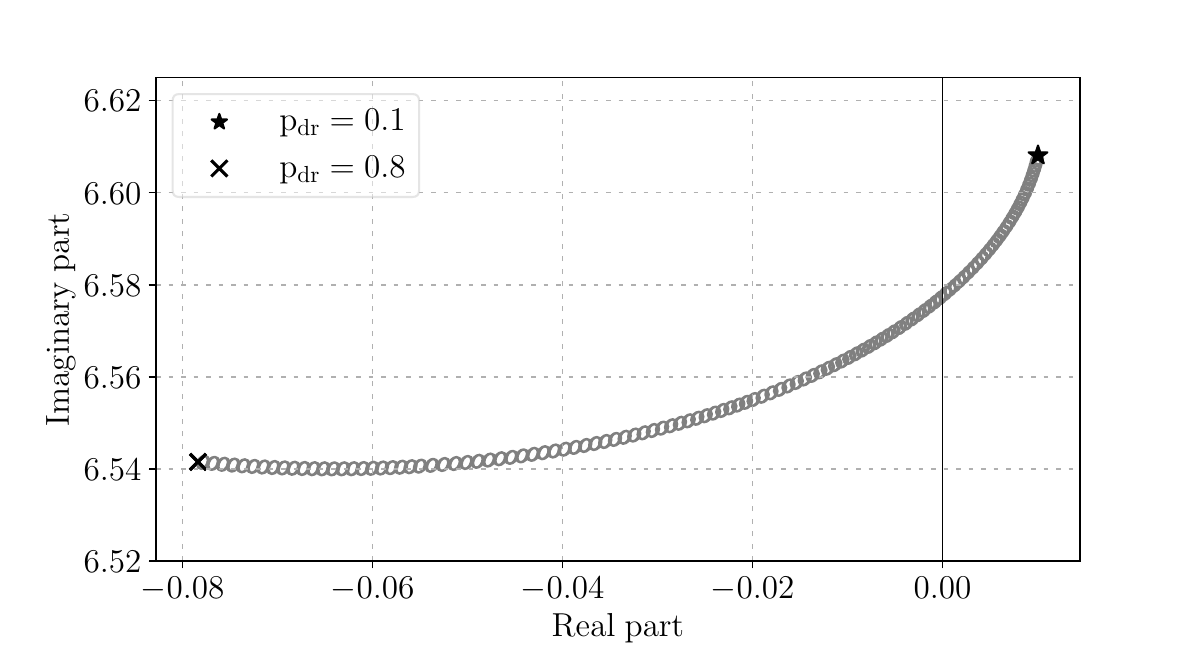}
    \caption{
    Tracing the interarea mode as $\rm p_{dr}$ is increased.}
    \label{fig:mod39b_wam_pdr_track}
  \end{subfigure}
  \hfill
  \begin{subfigure}{\columnwidth}
    \centering    \includegraphics[width=\linewidth]{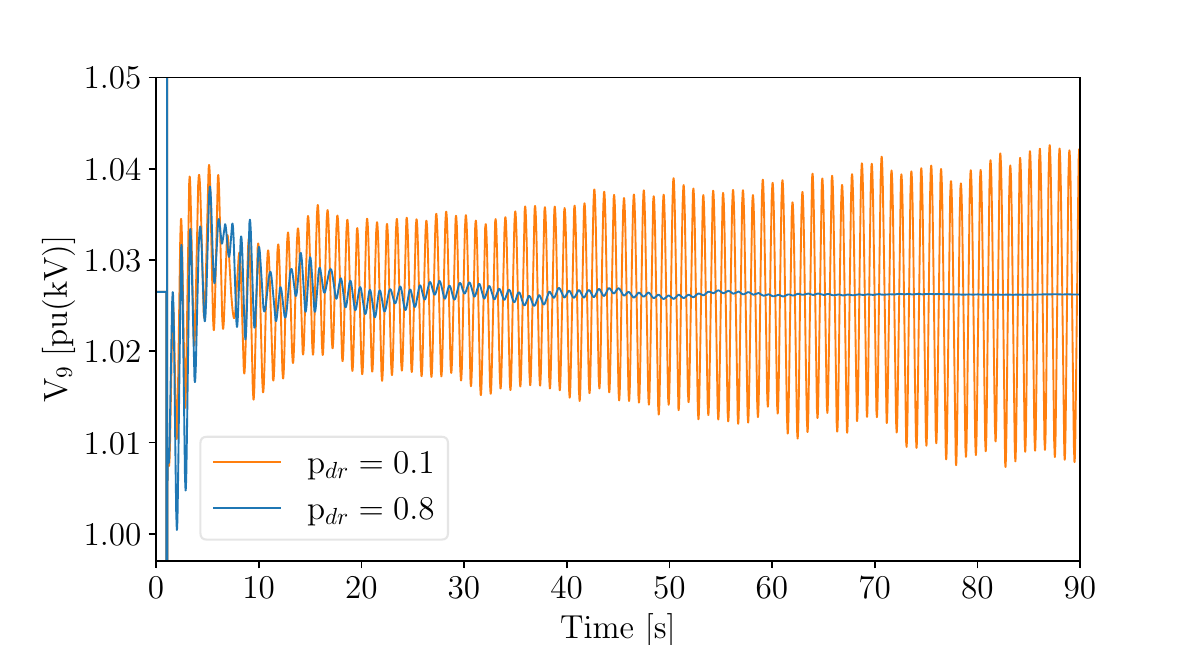}
    \caption{
    Time-domain simulation for different $\rm p_{dr}$ values.}
\label{fig:wams_tds_pdr}
  \end{subfigure}
  \caption{39-bus system: effect of increasing $\rm p_{dr}$ from $0.1$ to $0.8$.}
  \label{fig:mod39b_wam_pdr}
\end{figure}

Then we study the effect of increasing $p_{dr}$, when the normal delivery period is set to $0.09$~s. 
The resulting trajectory of the interarea mode is shown in Fig.~\ref{fig:mod39b_wam_pdr_track}. Increasing $p_{dr}$ has an effect similar to increasing $T$, driving the mode back into the stable region. 
This is consistent with both changes increasing the effective \ac{wams} latency and producing a trend similar to that observed in Fig.~\ref{fig:mod39b_PSS2_tau}.
Time-domain simulation of the same three-phase fault at bus~6 further validates this behavior (Fig.~\ref{fig:wams_tds_pdr}).
\color{black}

\subsection{Irish System}

\begin{figure}[pos=ht]
  \centering \includegraphics[width=\linewidth]{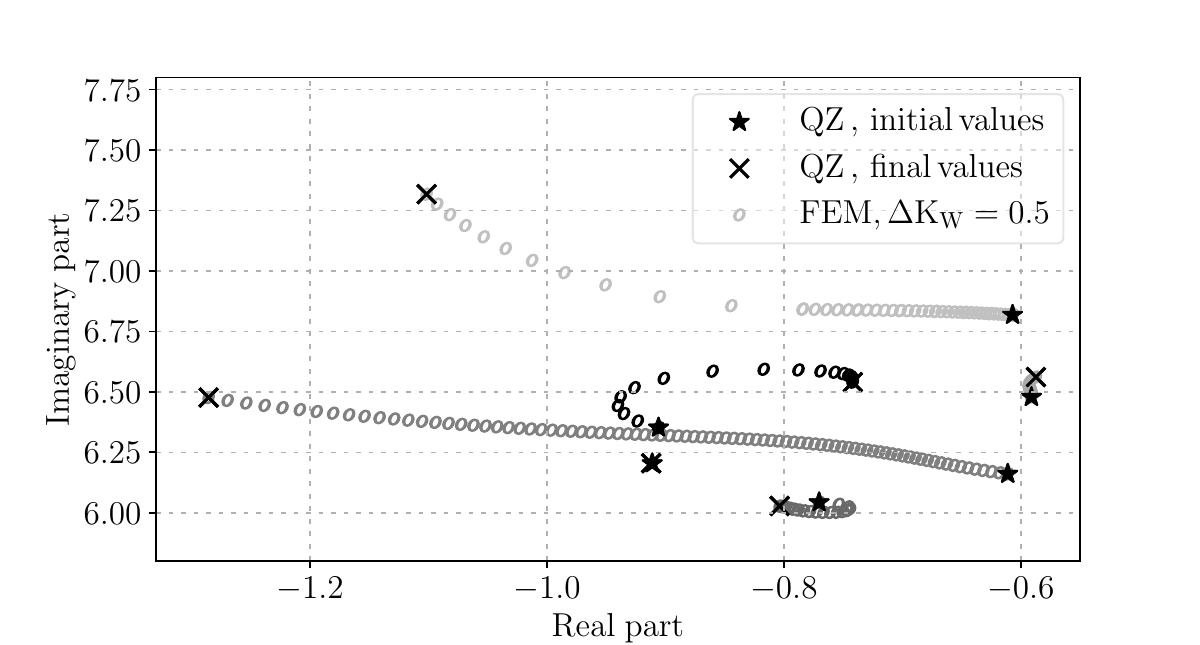}
  \caption{AIITS: tracking poorly damped electromechanical modes in the presence of multiple delays.}  \label{fig:AIITS_MD_D28_Kw_PSS_multiple}
\end{figure}

We assess the scalability of the proposed eigenvalue tracking method using a 1,502-bus dynamic model of the \acf{aiits}.  The delay-free \ac{dae} model has 1,629 states and 9,897 algebraic variables.  Its dimension, 11,526$\times$11,526, makes repeated use of standard dense QZ-based eigensolvers impractical.
We first examine the accuracy of \eqref{eq:sin_ret_singlevector:sys} in tracing poorly damped electromechanical modes of the \ac{aiits}.  The \ac{ddae} model includes 28 delayed variables, affecting voltage and frequency measurements of the system's \acp{avr} and \acp{pss}.  The delays, ranging from $0.009$ to $0.022$~s, increase the dimension of
the approximated linear eigenvalue problem to 11,834$\times$11,834.  
The effect of increasing the gains of the system's \acp{pss} from $1.5$ to $10$ is shown in Fig.~\ref{fig:AIITS_MD_D28_Kw_PSS_multiple}. Although the examined eigenvalues are tightly clustered in the complex plane, the proposed approach efficiently traces all of their trajectories.

\begin{figure}[pos=ht]
  \centering 
  \includegraphics[width=\linewidth]{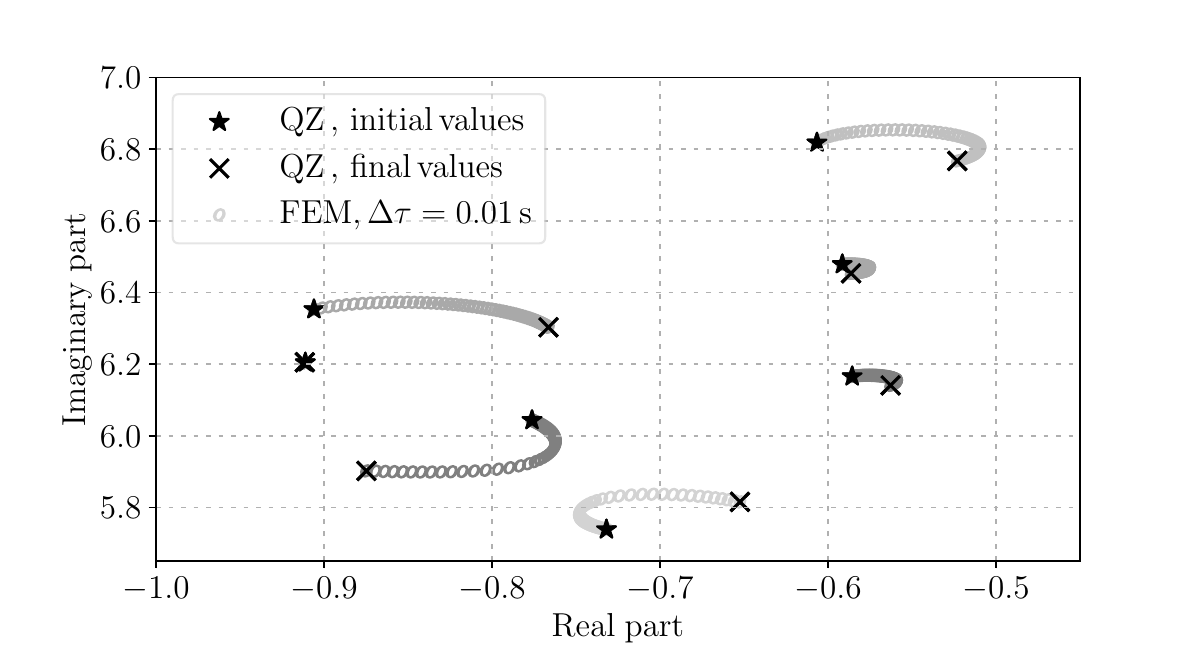}
  \caption{AIITS, SG at bus~717: increasing PSS input signal delay.}
  \label{fig:aiits_PSS2_tau}
\end{figure}

We next apply the formulation in Section~\ref{sec:trackdel} to trace the trajectories of the system's poorly damped electromechanical modes as the delay of the frequency signal of the \ac{pss} connected to the \ac{sg} at bus~717 increases from $0.01$ to $0.5$~s. 
The results, displayed in Fig.~\ref{fig:aiits_PSS2_tau}, confirm the accuracy of the proposed approach.

Lastly, we assess the computational cost of the proposed tracking approach, which can be broken down into the initial eigenvalue calculation and the continuation along the parameter trajectory. In this paper, the initial calculation is performed using the Chebyshev discretization described in \cite{Li_discr}. For $\mu$ delayed variables and $N$ Chebyshev nodes, this results in a generalized eigenvalue problem of dimension $d = \nxy + \mu (N+1)$.
For an eigensolver searching for $k \leq d$ eigenvalues, a rough estimate of the computational cost is 
$O(k^2 d + k^3)$, 
with a corresponding memory requirement of $O(k d + k^2)$. 
Practical algorithms 
may use a search subspace larger than $k$, so these estimates are approximate. For a full-spectrum dense eigensolution, $k=d$, and the above estimates become $O(d^3)$ for computational cost and $O(d
^2)$ for memory.

The continuation equations described by \eqref{eq:sin_ret_singlevector:sys} have dimension $2\nxy+2$. At each parameter step, the computational cost is dominated by sparse matrix factorizations and linear solves of this dimension.  The exact number of such operations depends on the numerical integration scheme and the number of Newton corrector iterations.  For example, employing the \ac{fem} requires one sparse factorization and linear solve per step, while each Newton corrector iteration introduces an additional factorization and solve. 
For a continuation matrix $\bfb M$ with LU factors $\bfb L$, $\bfb U$, the corresponding memory requirement is $O({\rm{nnz}}(\bfb M) +{\rm{nnz}}(\bfb L,\bfb U) + \nxy)$, where $\rm{nnz}(\cdot)$ denotes the number of nonzero matrix entries. The computational cost of the sparse factorizations and linear solves depends on the matrix sparsity pattern and ordering, and 
cannot, in general, be characterized solely as a function of the problem dimension.

We compare the computational speed of the proposed tracking approach with the repeated use of two eigendecomposition methods, QZ and \ac{ks}, which are suited to dense and sparse problems, respectively.
The comparison considers the case in which the time delay of the frequency signal of the PSS at bus 717 increases from 0.01 to 0.5 s over 49 parameter points.  The calculations are repeated over 10 runs, and the resulting mean computational times and standard deviations are summarized in Table~\ref{tab:aiits:fixed}\footnote{
Simulations are executed using a computer with a 13th Gen Intel(R) Xeon(R) Gold 5416S, 2.0 GHz (up to 3.8 GHz turbo) processor, 384~GB of RAM, and running a 64-bit Linux OS.}. 
The results show that the computational times of the three methods differ by orders of magnitude.  The proposed approach is substantially faster, while maintaining a strict error tolerance of $10^{-8}$ throughout the tracking.
\color{black}

\begin{table}[!ht]
\caption{
}
\label{tab:aiits:fixed}
\centering
\setlength{\tabcolsep}{4pt}
\renewcommand{\tabularxcolumn}[1]{m{#1}}
\begin{tabularx}{\linewidth}{
>{\raggedright\arraybackslash}X
>{\centering\arraybackslash}X
>{\centering\arraybackslash}X
>{\centering\arraybackslash}X
>{\centering\arraybackslash}X}
\toprule
\centering
{} & {} & {} & \multicolumn{2}{c}{Proposed Method} \\
 & {QZ} & {\ac{ks}} & {(\ac{tm})} & 
{(\ac{fem})} 
\\
\midrule
{Mean} & $176,313.0$~s &
$265.44$~s & $12.66$~s & $11.34$~s \\
\midrule
{Standard Deviation} & $214.2$~s & $6.88$~s & $0.75$~s & $0.81$~s \\
\bottomrule
\end{tabularx}
\end{table}

Both QZ and \ac{ks} compute multiple eigenvalues at once. QZ computes the full spectrum of the discretized problem, whereas \ac{ks} calculates only a selected part of it. However, repeated QZ and \ac{ks} return unordered eigenpairs at each parameter value. Therefore, reconstructing eigenvalue trajectories requires an additional mode-matching step, which becomes nontrivial for large sparse systems with clustered modes or partial-spectrum computations. The total time reported for QZ and \ac{ks} therefore includes only the eigensolution step and excludes any subsequent mode-matching procedure required to connect eigenpairs across parameter points.
In addition, the accuracy and the number of eigenvalues found by \ac{ks} depend strongly on the tuning of the \textit{shift-and-invert} transformation.  
In our case, for the values shown in Table~\ref{tab:aiits:fixed}, the number of iterations (on average $3$-$4$) and the number of computed eigenvalues (on average 633) vary across continuation steps.
The computational time of \ac{ks} can be reduced by targeting a smaller portion of the spectrum; however, selecting the spectral transformation and target region is nontrivial and typically requires trial and error, without guaranteeing a prescribed tolerance or the identification of the desired eigenvalue. In the absence of an efficient strategy to determine the transformation's spectral shift, it is kept constant at $\sigma=-0.15$. 
\color{black}

The proposed approach determines the trajectory of the targeted eigenpair in an automated manner and in a small fraction of the time required by the other methods. 
In the case considered in Table~\ref{tab:aiits:fixed}, 
the higher computational cost of the \ac{tm} is largely offset by its more accurate initial estimation, which can reduce the total number of Newton corrector iterations required for the prescribed error tolerance.
As a result, the \ac{fem} and the \ac{tm} exhibit similar computational times.
\color{black}

\section{Conclusion}
\label{sec:conclusions}

This paper proposes a continuation-based eigenvalue tracking technique for power systems impacted by time delays.  The formulation retains the sparsity of the system's \ac{ddae} model and uses a continuation parameter that allows multiple system properties, including delay magnitudes, to be expressed as functions of it and varied simultaneously. 
Case studies demonstrate the accuracy and computational efficiency of the proposed approach. 
Future work will extend the proposed method to track multiple eigenpairs simultaneously.
Additionally, the proposed approach will be applied to inverter-dominated power systems to track dynamic interactions in low-inertia grids and guide the design of delay-robust control schemes for \acp{der}.
\color{black}


\appendix
\section{Appendix}

In this section, we derive the expressions of $\bfb M$ and $\bfb h$ for the cases considered in Sections~\ref{sec:trackmul}, \ref{sec:trackdel},~\ref{sec:wams_del}, i.e., for systems with multiple delays, systems including a varying delay and systems including a \ac{wams} delay.

\subsection[Derivation]{Systems with Multiple Delays}
\label{deriv:mul_del}

We define the functions $h_r(t)=e^{-s_{\rm r} t}\cos{(s_{\rm i} t)}$ and $h_i(t)=e^{-s_{\rm r} t}\sin{(s_{\rm i} t)}$ and consider the notation $h^\del_r = h_r(\tau_\del)$ and $h^\del_i = h_i(\tau_\del)$.
By splitting real and imaginary parts of \eqref{eq:mul_ret_gep:diff3} and 
setting $\Cs = [ \bfb A_\del h^\del_r ]\T$, $\Sn  = [ \bfb A_\del h^\del_i]\T$, $\Csd = [ \bfb A'_\del h^\del_r ]\T$, $\Snd = [ \bfb A'_\del h^\del_i ]\T$, 
%
we obtain a system of the form \eqref{eq:sin_ret_singlevector:sys},
where
\begin{align*}
      \bfb h
     &= 
    \begin{bmatrix}
        (-s_{\rm r}\bfb{E}'+ \bfb{A}'_0 + \bfb J_\mu \Csd)\bfb\phi_{\rm r} + (s_{\rm i}\bfb{E}'+ \bfb J_\mu \Snd)\bfb\phi_{\rm i}\\
        (-s_{\rm i}\bfb{E}' - \bfb J_\mu \Snd)\bfb\phi_{\rm r} + (-s_{\rm r}\bfb{E}'+ \bfb{A}'_0 + \bfb J_\mu \Csd)\bfb\phi_{\rm i}\\ 
        0
        \\ 
        0
    \end{bmatrix}
\end{align*}

$\bfb M_{3}$ is the same as in \eqref{eq:sin_ret_singlevector:sys} and:

\begin{equation*}
  \begin{aligned}
    \bfb M_{1} &= 
    \begin{bmatrix}
        s_{\rm r}\bfb{E} - \bfb{A}_0 - \bfb J_\mu \Cs & -s_{\rm i} \bfb{E} - \bfb J_\mu \Sn  \\
        s_{\rm i} \bfb{E} + \bfb J_\mu \Sn  & s_{\rm r}\bfb{E} - \bfb{A}_0 - \bfb J_\mu \Cs
    \end{bmatrix}
\\
    \bfb M_{2} \! &= 
    \begin{bmatrix}
      \bfb{E}\bfb\phi_{\rm r} \!+\! \bfb J_{\tau \mu} ( \Cs \bfb\phi_{\rm r} + \Sn \bfb\phi_{\rm i}) 
      & \!\! -\bfb{E}\bfb\phi_{\rm i} \!-\! \bfb J_{\tau \mu}( \Cs \bfb\phi_{\rm r} \!-\! \Sn \bfb\phi_{\rm i})  \!
      \\
      \bfb{E}\bfb\phi_{\rm i} \!+\! \bfb J_{\tau \mu}( \Cs \bfb\phi_{\rm r} \!-\! \Sn \bfb\phi_{\rm i}) 
      & \!\! \bfb{E}\bfb\phi_{\rm r} \!+\! \bfb J_{\tau \mu}( \Cs \bfb\phi_{\rm r} \!+\! \Sn \bfb\phi_{\rm i}) 
    \end{bmatrix}
 \end{aligned}
\end{equation*}
with $\bfb J_\mu =  [\bfg{I}_r \; \bfg{I}_r \;\ldots \; \bfg{I}_r] \in \mathbb{R}^{r \times \mu r}$ and $\bfb J_{\tau \mu} = [\tau_1\bfg{I}_r \; \tau_2\bfg{I}_r \; \ldots \; \tau_\mu \bfg{I}_r] \in \mathbb{R}^{r \times \mu r}$; $\bfg{I}_r$ is the $r \times r$ identity matrix.

\subsection[Derivation]{Systems with Delay as Varying Parameter}
\label{deriv:sin_tau}

Splitting real and imaginary parts of \eqref{eq:sin_ret_gep:tau:der} 
and considering the functions $h_r$ and $h_i$ defined above for $\tau_\ell=p$, we arrive at the form of \eqref{eq:sin_ret_singlevector:sys},
where in this case:
\begin{align}
\label{eq:trackdel:h}
\bfb h
= 
\begin{bmatrix}
    \bfb{h}_1 \bfb\phi_{\rm r} +
    \bfb{h}_2 \bfb\phi_{\rm i}\\
    -\bfb{h}_2 \bfb\phi_{\rm r} + \bfb{h}_1\bfb\phi_{\rm i}\\ 
    0
    \\ 
    0
\end{bmatrix}
\end{align}
with
%
\begin{equation*}
  \begin{aligned}
    \;\;\, \bfb h_1 =& \;
        - \bfb A_\ell h^\ell_r s_r - \bfb A_\ell h^\ell_i s_i \\
    \;\;\, \bfb h_2 =& \;
        \bfb A_\ell h^\ell_r s_i - \bfb A_\ell h^\ell_i s_r 
  \end{aligned}
\end{equation*}
$\bfb M_{3}$ is the same as in \eqref{eq:sin_ret_singlevector:sys} and:
\begin{equation*}
  \begin{aligned}
    \bfb M_{1} \!= \!\!
    &
    \begin{bmatrix}
        s_{\rm r}\bfb{E}-\bfb{A}_0-\bfb A_\ell h^\ell_r & -s_{\rm i} \bfb{E} -\bfb A_\ell h^\ell_i \\
        s_{\rm i} \bfb{E} + \bfb A_\ell h^\ell_i & s_{\rm r}\bfb{E}-\bfb{A}_0-\bfb A_\ell h^\ell_r
    \end{bmatrix}
\\
    \bfb M_{2}\!=\!\!
    &
    \begin{bmatrix}
      \bfb{E} \bfb\phi_{\rm r} \!+\!  p \bfb A_\ell  (\bfb\phi_{\rm r} h^\ell_r \!+\! \bfb\phi_{\rm i}  h^\ell_i) 
      & \!\!-\bfb{E} \bfb\phi_{\rm i} \!+\! p \bfb A_\ell  (\bfb\phi_{\rm r} h^\ell_i \!-\! \bfb\phi_{\rm i}  h^\ell_r) 
      \!
      \\ \!
      \bfb{E} \bfb\phi_{\rm i} \!-\!  p \bfb A_\ell  (\bfb\phi_{\rm r} h^\ell_i \!-\!  \bfb\phi_{\rm i}  h^\ell_r) \!
      & \! \bfb{E} \bfb\phi_{\rm r} \!+\!  p \bfb A_\ell  (\bfb\phi_{\rm r} h^\ell_r \!+\! \bfb\phi_{\rm i}  h^\ell_i) 
      \!
    \end{bmatrix}
\end{aligned}
\end{equation*}

If the system includes multiple delays, then 
the matrices $\Cs $, $\Sn $, $\bfb J_\mu$ and $\bfb J_{\tau \mu}$ are also considered.
Splitting real and imaginary parts of \eqref{eq:mul_tau_ret_gep} 
we arrive at the form of  \eqref{eq:sin_ret_singlevector:sys}, 
with $\bfb h$ given by \eqref{eq:trackdel:h} and where in this case 
\begin{equation*}
\begin{aligned}
\bfb h_1 =&
    - \bfb A_\ell h^\ell_r s_r - \bfb A_\ell h^\ell_i s_i\\
\bfb h_2 =&
    \bfb A_\ell h^\ell_r s_i - \bfb A_\ell h^\ell_i s_r
\end{aligned}
\end{equation*}
$\bfb M_{3}$ is the same as in \eqref{eq:sin_ret_singlevector:sys} and:
\begin{equation*}
  \begin{aligned}
    \bfb M_{1} \!&=\! 
    \begin{bmatrix}
        s_{\rm r}\bfb{E}-\bfb{A}_0 \!-\! \bfb A_\ell h^\ell_r\! - \! \bfb J_\mu \Cs \!\!&\!\! -s_{\rm i} \bfb{E} -\bfb A_\ell h^\ell_i - \bfb J_\mu \Sn  \\
        s_{\rm i} \bfb{E} + \bfb A_\ell h^\ell_i + \bfb J_\mu \Sn  \!\!&\!\! s_{\rm r}\bfb{E} \!-\! \bfb{A}_0 \!-\! \bfb A_\ell h^\ell_r - \bfb J_\mu \Cs 
    \end{bmatrix}
\\
    \bfb M_{2} &=
    \left[
    \begin{matrix}
      \bfb{E}\bfb\phi_{\rm r} + \bfb J_{\tau \mu} ( \Cs \bfb\phi_{\rm r} + \Sn \bfb\phi_{\rm i}) + \bfb A_\ell  p  (\bfb\phi_{\rm r} h^\ell_r +\bfb\phi_{\rm i}  h^\ell_i)
      & \!\!\!\!\ldots \\
      \bfb{E}\bfb\phi_{\rm i} + \bfb J_{\tau \mu}( \Cs \bfb\phi_{\rm i} - \Sn \bfb\phi_{\rm r})  - \bfb A_\ell  p  (\bfb\phi_{\rm r} h^\ell_i -  \bfb\phi_{\rm i}  h^\ell_r)
      & \!\!\!\!\ldots
    \end{matrix}
    \right.\\
    &\left.
    \begin{matrix}
       \ldots \!\!\! &-\bfb{E}\bfb\phi_{\rm i} - \bfb J_{\tau \mu}( \Cs \bfb\phi_{\rm i} - \Sn \bfb\phi_{\rm r})  + \bfb A_\ell  p  (\bfb\phi_{\rm r} h^\ell_i -  \bfb\phi_{\rm i}  h^\ell_r) \\
       \ldots \!\!\!&\bfb{E}\bfb\phi_{\rm r} + \bfb J_{\tau \mu} ( \Cs \bfb\phi_{\rm r} + \Sn  \bfb\phi_{\rm i})  + \bfb A_\ell  p  (\bfb\phi_{\rm r} h^\ell_r +\bfb\phi_{\rm i}  h^\ell_i)
    \end{matrix}
    \right]
  \end{aligned}
\end{equation*}

\subsection[Derivation]{Communication Delays with Noise and Data Dropouts}
\label{deriv:wams_del}

We set ${\rm S}_T^r=\Re\{{\rm S}_T\}$, ${\rm S}_T^i=\Im\{{\rm S}_T\}$, $\tilde{\bfb S}_{TD}^r=\Re\{{\rm S}_{TD}\} \bfb{A}_1$, $\tilde{\bfb S}_{TD}^i=\Im\{{\rm S}_{TD}\} \bfb{A}_1$, 
${\rm S}_{Tp}^r=\Re\{{\rm S}_{Tp}\}$ and ${\rm S}_{Tp}^i=\Im\{{\rm S}_{Tp}\}$. 
\color{black} 
Splitting real and imaginary parts of \eqref{eq:wams_diff} leads to a system in the form of \eqref{eq:sin_ret_singlevector:sys}, where in this case $\bfb M_{3}$ is the same as in \eqref{eq:sin_ret_singlevector:sys} and:
\begin{equation*}
\begin{aligned}
\bfb h
    & = 
    \begin{bmatrix}
        \bfb{h}_1\\
        \bfb{h}_2\\ 
        0
        \\ 
        0
    \end{bmatrix} \\
    \color{black}
\bfb M_{1} &=
    \begin{bmatrix}
        s_{\rm r}\bfb{E}-\bfb{A}_0 - {\rm S}_T^r \bfb{A}_1 & -s_{\rm i} \bfb{E} + {\rm S}_T^i  \bfb{A}_1\\
        s_{\rm i} \bfb{E} - {\rm S}_T^i  \bfb{A}_1  & s_{\rm r}\bfb{E} - \bfb{A}_0 - {\rm S}_T^r \bfb{A}_1
\end{bmatrix}
\\
    \bfb M_{2} &= 
    \begin{bmatrix}
      (\bfb{E}\! - \tilde{\bfb S}_{TD}^r) \bfb\phi_{\rm r}\! +\! \tilde{\bfb S}_{TD}^i \bfb\phi_{\rm i} 
      &\!\! -(\bfb{E}\! -\! \tilde{\bfb S}_{TD}^r) \bfb\phi_{\rm i}\! +\! \tilde{\bfb S}_{TD}^i  \bfb\phi_{\rm r} \\
      (\bfb{E}\! -\! \tilde{\bfb S}_{TD}^r) \bfb\phi_{\rm i}\! -\! \tilde{\bfb S}_{TD}^i  \bfb\phi_{\rm r} 
      &\!\! (\bfb{E}\! - \tilde{\bfb S}_{TD}^r) \bfb\phi_{\rm r}\! +\! \tilde{\bfb S}_{TD}^i  \bfb\phi_{\rm i}
    \end{bmatrix} 
  \end{aligned}
\end{equation*}
%
%
where $\bfb{h}_1=(-s_{\rm r}\bfb{E}'+ \bfb A'_0 + {\rm S}_{Tp}^r\bfb A_1 + {\rm S}_{T}^r\bfb A'_1)\bfb\phi_{\rm r} + (s_{\rm i}\bfb{E}' - {\rm S}_{Tp}^i \bfb A_1 - {\rm S}_{T}^i \bfb A'_1)\bfb\phi_{\rm i}$ and $\bfb{h}_2=- (s_{\rm i}\bfb{E}' - {\rm S}_{Tp}^i \bfb A_1 - {\rm S}_{T}^i \bfb A'_1)\bfb\phi_{\rm r} + (-s_{\rm r}\bfb{E}'+ \bfb A'_0 + {\rm S}_{Tp}^r\bfb A_1 + {\rm S}_{T}^r\bfb A'_1)\bfb\phi_{\rm i}$.

We note that when the continuation parameter affects only the functions $h_p$ and $h_s$, i.e., \ac{wams} delay parameters such as $T$ or $p_{dr}$, the matrices $\bfb A$ and $\bfb E$ do not depend on $p$, and therefore:
\begin{equation*}
    \bfb A_0'=\bfb A_1'=\bfb E'= \bfb 0 
\end{equation*}
Conversely, if the continuation parameter is independent of the \ac{wams} delay parameters, then $S_{Tp}=0$. In both the above cases, \eqref{eq:wams_diff} can be simplified. 

\color{black}


\printcredits

\bibliographystyle{IEEEtran}
\bibliography{refs_unmarked}





\end{document}